\documentclass[10pt,journal]{IEEEtran}
\ifCLASSINFOpdf
\else
\fi
\usepackage{times,amsmath,epsfig}
\usepackage[latin1]{inputenc}
\usepackage{fancyhdr}
\usepackage{amsmath}
\usepackage{amsfonts}
\usepackage{amssymb}
\usepackage{float}
\usepackage{array}
\usepackage{graphicx}
\usepackage{mathtools}
\graphicspath{{./fig/}}
\usepackage{epstopdf}
\usepackage{url}
\usepackage{subfigure}
\usepackage{bm}
\usepackage{breqn}
\usepackage{xcolor}
\usepackage{soul}
\usepackage{picinpar,graphicx}
\usepackage{cite}
\usepackage{multirow}
\usepackage{geometry}
\newgeometry{top=2cm,left=1.778cm,right=1.778cm,bottom=2cm}
\usepackage{algorithm}
\usepackage{algpseudocode}

\IEEEoverridecommandlockouts
\graphicspath{{Figures_eps/}}

\begin{document}
\title{Cooperative Sensing-Assisted Predictive Beam Tracking for MIMO-OFDM Networked ISAC Systems}
\author{Xiaoyu Yang, \IEEEmembership{Graduate Student Member, IEEE}, Zhiqing Wei, \IEEEmembership{Member, IEEE}, Jie Xu, \IEEEmembership{Fellow, IEEE}, Huici Wu, \IEEEmembership{Member, IEEE}, and Zhiyong Feng, \IEEEmembership{Senior Member, IEEE}
\thanks{Xiaoyu Yang, Zhiqing Wei, and Zhiyong Feng are with the Key Laboratory of Universal Wireless Communications, Ministry of Education, Beijing University of Posts and Telecommunications, Beijing 100876, China (e-mail: \{xiaoyu.yang, weizhiqing, fengzy\}@bupt.edu.cn). }
\thanks{Jie Xu is with the School of Science and Engineering (SSE), the Shenzhen Future Network of Intelligence Institute (FNii-Shenzhen), and the Guangdong Provincial Key Laboratory of Future Networks of Intelligence, The Chinese University of Hong Kong (Shenzhen), Guangdong 518172, China (e-mail: xujie@cuhk.edu.cn).}
\thanks{Huici Wu is with the National Engineering Research Center of Mobile Network Technologies, Beijing University of Posts and Telecommunications,
Beijing 100876, China, and also with the Pengcheng Laboratory, Shenzhen 518066, China (e-mail: dailywu@bupt.edu.cn).}
}

\maketitle
\begin{abstract}
This paper studies a multiple-input multiple-output (MIMO) orthogonal frequency division multiplexing (OFDM) networked integrated sensing and communication (ISAC) system, in which multiple base stations (BSs) perform beam tracking to communicate with a mobile device.
In particular, we focus on the beam tracking over a number of tracking time slots (TTSs) and suppose that these BSs operate at non-overlapping frequency bands to avoid the severe inter-cell interference.
Under this setup, we propose a new cooperative sensing-assisted predictive beam tracking design. In each TTS, the BSs use echo signals to cooperatively track the mobile device as a sensing target, and continuously adjust the beam directions to follow the device for enhancing the performance for both communication and sensing.
First, we propose a cooperative sensing design to track the device, in which the BSs first employ the two-dimensional discrete Fourier transform (2D-DFT) technique to perform local target estimation, and then use the extended Kalman filter (EKF) method to fuse their individual measurement results for predicting the target parameters.
Next, based on the predicted results, we obtain the achievable rate for communication and the predicted conditional Cram\'er-Rao lower bound (PC-CRLB) for target parameters estimation in the next TTS, as a function of the beamforming vectors.
Accordingly, we formulate the predictive beamforming design problem, with the objective of maximizing the achievable communication rate in the following TTS, while satisfying the PC-CRLB requirement for sensing.
To address the resulting non-convex problem, we first propose a semi-definite relaxation (SDR)-based algorithm to obtain the optimal solution, and then develop an alternative penalty-based algorithm to  get a high-quality low-complexity solution.
Simulation results indicate that the proposed cooperative sensing design achieves higher target tracking accuracy than other benchmark schemes.
The results also validate the benefits of multi-BS cooperative sensing in improving tracking performance compared with the conventional single-BS sensing.
\end{abstract}


\IEEEpeerreviewmaketitle
\section{Introduction}
\subsection{Background}
Integrated sensing and communication (ISAC) has been envisioned as a promising technology for sixth-generation (6G) wireless networks, which possesses the capability to support a variety of emerging applications such as smart manufacturing, autonomous driving, and unmanned aerial vehicle (UAV) navigation \cite{feng2020joint}. In specific, ISAC enables the integration of communication and sensing functionalities on a unified platform by sharing hardware architectures, signal processing modules, as well as scarce spectrum and energy resources.
Based on this shared configuration, ISAC is able to enhance the spectrum and energy efficiency, as well as reduce the hardware costs and energy consumption \cite{liu2022integrated}. In addition, the integration of communication and sensing enables their mutual enhancements. On the one hand, ISAC systems can leverage sensing to acquire propagation environment information and accordingly adjust the communication parameters dynamically, thereby significantly improving communication quality and robustness (see, e.g., \cite{zeng2024tutorial,ren2024sensing}).
On the other hand, ISAC systems can exploit existing networked communication infrastructures to achieve high-speed and reliable transmission of sensing information, thus enhancing sensing coverage and accuracy (see, e.g., \cite{10273396,10226276}).

Orthogonal frequency division multiplexing (OFDM) is expected to be widely used in 6G wireless networks for both communication and sensing \cite{solaija2024orthogonal}.
From the wireless communication perspective, OFDM waveforms offer advantages such as resilience to frequency-selective fading and high spectral efficiency. From the wireless sensing perspective, OFDM waveforms allow for the decoupling of time delay and Doppler measurements, thus achieving superior target estimation performance as compared to the conventional radar waveforms. Therefore, it is practically attractive to implement OFDM waveforms for ISAC.
On the other hand, multiple-input multiple-output (MIMO) has emerged as a promising technology to enhance the performance of both communication and radar systems \cite{10480333}, which offers new spatial degrees of freedom (DoFs) to provide beamforming, spatial multiplexing, and spatial diversity gains.
To reap both benefits of OFDM and MIMO, it is becoming increasingly important to investigate the design and optimization of MIMO-OFDM ISAC systems (see, e.g., \cite{keskin2021mimo}), which is the main focus of this paper.

\subsection{Related Works}
Efficient beamforming design is crucial for the success of MIMO-OFDM ISAC systems. This, however, is a difficult task in practice, which requires the base station (BS) to obtain the accurate pointing direction or channel information of mobile devices. Among various beamforming approaches, beam training is widely adopted, in which the BS transmits training beams to scan the entire angular space and the device feeds back the signal strength of each beam. This enables the BS to identify the optimal beam pair for the device.
However, beam training is applicable for static or slowly-changing scenarios, as significant overheads may be induced in dynamic scenarios such as autonomous driving and UAVs. To deal with this issue, beam tracking becomes necessary.
In conventional beam tracking \cite{liu2019ekf}, the BS sends pilots to the mobile device, which estimates the angle and feeds it back. Nevertheless, such feedback-based scheme results in considerable latency inevitably, which may compromise the performance seriously.
Alternatively, predictive beam tracking has emerged as a promising solution to reduce the tracking latency by predicting the device position in advance to facilitate the beamforming design.
In addition, the advancements of ISAC enable the exploitation of sensing to enhance predictive beam tracking by leveraging echo signals to acquire the state of the mobile device, eliminating the feedback process.
Furthermore, unlike conventional beam tracking that employs a limited number of pilots, the ISAC-enabled scheme leverages the entire downlink data block for communication and sensing. Although the echo signal power is attenuated due to round-trip path loss, utilizing the signal from the entire data block allows for energy accumulation, improving the received signal-to-noise ratio (SNR) and avoiding additional training or pilot overheads.

In the literature, some initial efforts investigated the ISAC-enabled predictive beam tracking \cite{liu2020radar,yuan2020bayesian,zhang2024predictive,9947033,cui2023seeing}. For instance, Liu \emph{et al.} in \cite{liu2020radar} proposed a sensing-aided predictive beamforming method for vehicle-to-infrastructure (V2I) communication, in which the BS processes echo signals to estimate parameters of vehicles and then uses the Extended Kalman Filter (EKF) method to predict the angles of vehicles for beam alignment.
To reduce the signaling overhead, Yuan \emph{et al.} in \cite{yuan2020bayesian} proposed a message passing (MP) method to predict the angles of vehicles, with beamforming designed to align with the predicted angles.
Zhang \emph{et al.} in \cite{zhang2024predictive} proposed a deep learning (DL)-based framework to facilitate predictive beamforming design by bypassing explicit channel prediction.
Furthermore, Du \emph{et al.} in \cite{9947033} modeled the vehicle as an extended target, employed the EKF method to predict its angle, and proposed a dynamic predictive beam tracking method with varying beamwidths.
Moreover, considering more realistic environments with multipath channels, Cui \emph{et al.} in \cite{cui2023seeing} proposed an ISAC-aided EKF-based beam tracking method.
Despite the above progress on ISAC-based predictive beam tracking, these prior works focused on scenarios with a single BS, overlooking the promising cooperation gain of multiple densely deployed BSs in 6G. Furthermore, these studies considered simplified beam alignment towards the predicted target position, which cannot achieve the optimal overall performance for both communication and sensing.

Exploiting multi-BS cooperation for networked ISAC is an efficient way to improve both communication and sensing performance \cite{10273396}. On the one hand, multi-BS cooperative communication suppresses the inter-cell interference, thus improving achievable rates. On the other hand, multi-BS cooperative sensing provides richer target information from diverse observation angles, thereby enhancing sensing coverage, accuracy, and resolution.
Inspired by the above appealing advantages, Cheng \emph{et al.} in \cite{cheng2024optimal} studied a networked ISAC system for joint detection and communication, in which the transmit beamforming at the BSs were optimized to maximize  the detection probability, subject to the signal-to-interference-plus-noise ratio (SINR) requirements for communication. In \cite{huang2024edge}, Huang \emph{et al.} designed the sensing scheduling, together with the beamforming for sensing and beamforming for offloading transmission at multiple BSs, with the aim of minimizing the power consumption, while ensuring constraints on offloading throughput, beam pattern error, and mean-squared cross correlation pattern.
Besides, Li \emph{et al.} in \cite{li2023towards} studied the multi-BS beamforming optimization to maximize the sensing SNR, while ensuring the communication SINR constraints.
However, these existing works only investigated the beamforming optimization in static narrowband systems for fulfilling communication and sensing tasks at the same time. It is still unclear how to exploit the  symbiotic gain between communication and sensing (e.g., sensing-assisted communication of our interest) via multi-BS cooperation.

\subsection{Contributions}
Different from prior works, this paper studies the cooperative sensing-assisted predictive beam tracking for a mobile device/sensing target\footnote{In general, sensing targets can be classified into two categories: cooperative targets with communication requirements and non-cooperative targets without such requirements temporarily. To investigate the auxiliary gain of sensing on communication, without loss of generality, this work focuses on the cooperative target \cite{liu2020radar,yuan2020bayesian,zhang2024predictive,9947033,cui2023seeing}. Specifically, cooperative targets receive communication information to obtain the surrounding environmental information sensed by the BSs, such as obstacles, enabling them to efficiently and safely perform tasks through motion/trajectory control.} in MIMO-OFDM networked ISAC systems.
This introduces the following new technical challenges.
Firstly, different from the case of a single transmit antenna and/or narrowband transmission, in MIMO-OFDM systems, the transmitted signals over all subcarriers carrying random communication symbols would be coupled with target parameters in the spatial domain. This coupling effect makes it infeasible to directly separate random communication symbols and target parameters, thus greatly complicating the target estimation.
Next, multiple BSs obtains target estimation results from different observation angles. In this case, it is challenging to fuse these results to achieve efficient cooperative target sensing.
Furthermore, different from narrowband systems that only need to align the beam to the predicted target position, wideband OFDM systems need to optimize the predictive beamforming over multiple subcarriers, thus introducing additional design DoFs and increased implementation complexity.
As such, the resulting predictive beamforming optimization problem is difficult to tackle in the multi-BS wideband scenario.
This motivates this work to address these issues.

This paper studies a MIMO-OFDM networked ISAC system, in which multiple BSs employ beam tracking to communicate with a mobile device over a number of tracking time slots (TTSs), with predictive beamforming being implemented at each individual TTS. We propose a cooperative sensing-assisted predictive beam tracking scheme, in which these BSs dynamically adjust the beam directions to track the mobile device as a moving target utilizing the target parameters obtained by processing the received echo signals.
In particular, the main results of this paper are listed as follows.
\begin{itemize}
\item
First, we propose a cooperative sensing-based target tracking scheme. Specifically, we develop a local target estimation approach for multiple BSs to estimate the time delay and Doppler frequency shift
of the moving target based on the two-dimensional discrete Fourier transform (2D-DFT). Then, the EKF-based target tracking approach is proposed to fuse individual measurements from BSs for predicting the target position and velocity.
\item
Subsequently, we propose a predictive beamforming design scheme. In specific, based on the predicted target parameters, we obtain the corresponding achievable communication rates over all subcarriers and the predicted conditional Cram\'er-Rao lower bound (PC-CRLB) for sensing. Correspondingly, the multi-BS predictive beamforming design problem is formulated, with the objective of maximizing the achievable rate subject to the PC-CRLB requirement for target position estimation.
\item
To address the resulting problem, we first transform the PC-CRLB constraint into tractable forms by utilizing the Schur complement, and then propose a semi-definite relaxation (SDR)-based algorithm to obtain the optimal predictive beamforming solution. Then, to reduce computational complexity, we further develop a penalty-based algorithm to obtain a high-quality low-complexity solution by leveraging the successive convex approximation (SCA) technique.
\item
Finally, simulation results show that the proposed schemes significantly outperform the benchmark schemes in target tracking. Furthermore, it is shown that multi-BS cooperative sensing achieves better target tracking performance compared with the single-BS counterpart.
\end{itemize}

The remainder of this paper is organized as follows.
In Section II, we introduce the system model.
Section III proposes the cooperative sensing-assisted target racking scheme.
In Section IV, we formulate the predictive beamforming design problem, and propose an SDR-based algorithm and a penalty-based algorithm to solve it, respectively.
Section V presents numerical results to show the performance of the proposed schemes.
Finally, Section VI concludes this paper.

$\emph{Notations}$: In this paper, vectors and matrices are denoted by bold-face lower-case and upper-case letters, respectively.
$\mathbb{C}^{N \times M}$ denotes the set of $N \times M$ complex-valued matrices.
$\|\mathbf{X}\|$, $\mathbf{X}^{-1}$, and $\mathrm{Tr}(\mathbf{X})$ stand for the Euclidean norm, inverse, and trace of matrix $\mathbf{X}$, respectively.
Transpose, conjugate, and conjugate transpose operations are denoted by ${{\left( \cdot \right)}^{T}}$, ${{\left( \cdot \right)}^{*}}$, and ${{\left( \cdot  \right)}^{H}}$, respectively.
${\mathbf{{X}}}\underset{\scriptscriptstyle}{\succeq }\mathbf{0}$ means that $\mathbf{X}$ is a positive semidefinite matrix.
$\mathbb{E}\left\{ \cdot \right\}$ represents the statistical expectation.
$\otimes$ and $\odot$ stand for the Kronecker product and Hadamard product, respectively.
$\mathrm{diag}\left( \mathbf{x} \right)$ stands for a diagonal matrix, where each diagonal element is the corresponding element in $\mathbf{x}$.
$\mathrm{blkdiag}\left( \mathbf{X}_1,\cdots,\mathbf{X}_i\right)$ denotes a block diagonal matrix, where diagonal components are $\mathbf{X}_1,\cdots,\mathbf{X}_i$.
${\cal C}{\cal N}\left(\mathbf{x},\mathbf{Y}\right)$ represents the circularly symmetric complex Gaussian (CSCG) random vector with mean vector $\mathbf{x}$ and covariance matrix $\mathbf{Y}$.

\section{System Model}
\begin{figure}[t]
\noindent \centering{}\includegraphics[scale=0.54]{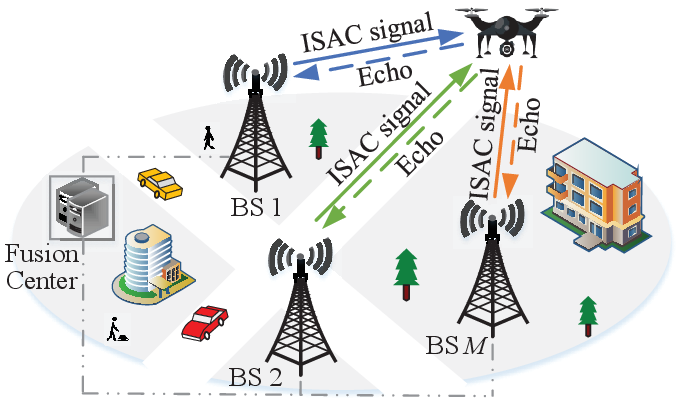}\caption{A networked ISAC system.}\label{Scene}
\end{figure}
As shown in Fig. \ref{Scene}, we consider a MIMO-OFDM networked ISAC system with $M$ BSs indexed by ${\cal M} = \left\{ {1, \cdots ,M} \right\}$, operating at the millimeter wave (mmWave) band to cooperatively track a single-antenna mobile device/sensing target.
As for sensing operations, the mobile device is regarded as a point-like target.
It is assumed that the BSs employ full-duplex operation with sufficiently isolated transmit and receive antennas to perform downlink communication while receiving echoes for sensing.
Each BS is equipped with a uniform linear array (ULA) of $N_t >1$ transmit antennas and $N_r >1$ receive antennas.
The target is observed over time interval $t \in \left[ {0,T} \right]$ with the maximum time duration of interest being $T$.
For ease of exposition, we divide the time duration $T$ into a number of TTSs each with duration $\Delta T$.
By assuming $\Delta T$ to be sufficiently small, the target parameters remain constant within each TTS.
Let ${\mathbf{b}_m} = {\left[ {b_x^m,b_y^m} \right]^T}$ and ${\mathbf{p}_n} = {\left[ {p_x^n,p_y^n} \right]^T}$ denote the positions of BS $m$ and the device at TTS $n$, respectively.

\subsection{Design Framework}
\subsubsection{Cooperation Among BSs}
To facilitate cooperation, all BSs are connected to a fusion center (FC) via fronthaul links for information sharing.
In fact, the fronthaul links are limited in capacity and latency, while the FC faces constraints in computational resources and processing efficiency. To characterize the performance limits, we assume that the fronthaul links and FC can sufficiently meet the exchange and processing requirements of the networked ISAC system.
Besides, the BSs are assumed to achieve perfect synchronization with the assistance of the FC.
It is known that the frequency division (FD) scheme\footnote{In particular, apart from the FD scheme, other techniques for separating signals from different BSs include time division (TD) and code division (CD) schemes. Although the TD scheme allows for better separation of transmit signals from BSs \cite{9528013}, it cannot facilitate cooperative sensing of the target. In contrast, the CD scheme enables cooperative sensing but involves complex coding design and signal processing to effectively distinguish the signals \cite{9359665}.} is a widely employed multi-cell frequency reuse scheme in mobile communication networks for distinguishing the transmit signals of different BSs \cite{10616023}.
Therefore, assuming sufficient spectrum resources, particularly in the considered high-frequency band, the BSs are assumed to operate at non-overlapping frequency bands to mitigate the adverse effects of inter-cell interference.
In the studied system, the BSs send individual signals to the device/target and receive the respective target-reflection signals for local signal processing. Then, the local measurement results are conveyed to the FC, enabling multi-BS cooperative sensing of the target through data-level fusion\footnote{Fusion strategies are generally divided into data-level fusion and signal-level fusion. In the data-level fusion, BSs perform individual signal processing and convey the results to the FC for data fusion \cite{9551293}; in the signal-level fusion, BSs directly convey the received echo signals to the FC for joint processing \cite{10226276}. However, signal-level fusion has higher requirements for the capacity and latency of the fronthaul links among BSs as well as their deployment locations. Hence, the focus of this paper is on the data-level fusion.}.
This cooperative configuration facilitates the sharing of target and system information, such as synchronization parameters, while reducing the capacity and latency requirements of the fronthaul links.

\subsubsection{Proposed Protocol}
\begin{figure*}
\noindent \centering{}\includegraphics[scale=0.4]{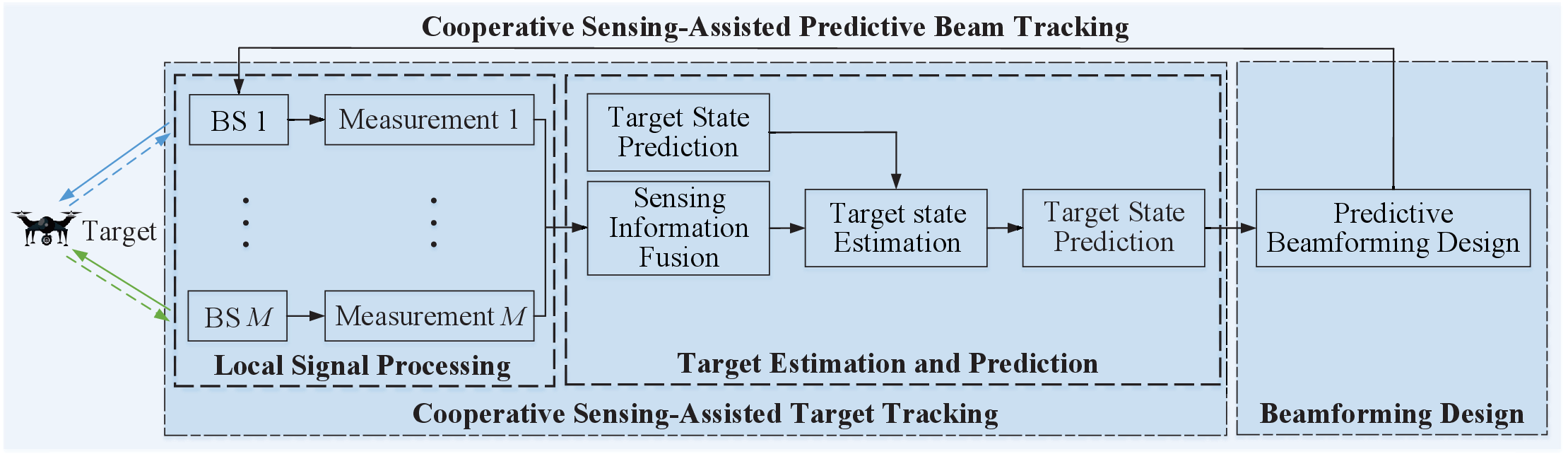}\caption{Two-stage design framework for predictive beam tracking.}\label{outline}
\end{figure*}
To improve tracking accuracy and communication quality with the mobile device/sensing target, we propose a cooperative sensing-assisted predictive beam tracking scheme. Specifically, as depicted in Fig. \ref{outline}, the proposed scheme is divided into two stages within each TTS, including stage I for cooperative sensing-assisted target tracking and stage II for predictive beamforming design, respectively.
In stage I, the BSs transmit individual signals steered by the beamformers designed by the FC at the previous TTS for both communication and sensing. After the round-trip time delay, the BSs receive the echo signals and perform local signal processing to estimate the device/target locally.
Since the BSs operate in full-duplex mode with sufficient separation of transmit and receive antennas, they can maintain communication with the device while receiving echo signals.
Next, the local measurement results are conveyed to the FC for data-level fusion, enabling target estimation through cooperative sensing.
Subsequently, based on this estimation, the FC employs the state evolution model to perform target prediction.
In stage II, leveraging the target prediction, the FC designs predictive beamforming vectors over all subcarriers for the next TTS\footnote{The proposed framework can be integrated with edge servers and the cloud radio access network (C-RAN) architecture to enhance the applicability of the developed schemes \cite{yan2024deep,huang2024edge}, with the related design left for future work.}.
The detailed designs for both stages are in Sections III and IV, respectively.

\subsection{Signal Model}
In the considered system, each BS $m$ is operated over a separate frequency band with central subcarrier frequency $f_m, m \in {\cal M}$ and bandwidth $B$.
For ease of representation, the frequency band for each BS is separated into $K$ subcarriers indexed by ${\cal K} = \left\{ {0, \cdots, K-1} \right\}$ each with bandwidth $\Delta f$.
Besides, each OFDM block consists of $L$ consecutive OFDM symbols each with duration $T_s$, where $T_s = T_e + T_{c}$ with $T_e$ denoting the elementary symbol duration and $T_{c}$ denoting the cyclic prefix (CP) duration.
Let $s_{m,k}^n\left[ l \right]$ denote the data symbol at the $k$-th subcarrier of the $l$-th OFDM symbol sent by BS $m$ at TTS $n$.
The data symbols are precoded by a beamforming vector $\mathbf{w}_{m,k}^n \in \mathbb{C}{^{{N_t} \times 1}}$ for subcarrier $k$ in frequency domain to focus the beam on the device. Hence, the transmitted baseband frequency-domain signal of BS $m$ at subcarrier $k$, OFDM symbol $l$, and TTS $n$ is $\mathbf{x}_{m,k}^n\left[ l \right] = \mathbf{w}_{m,k}^n s_{m,k}^n\left[ l \right]$.
Then, frequency-domain signals are transformed into temporal domain by the $K$-point inverse discrete Fourier transform (IDFT), and a CP with duration $T_{c}$ is added to avoid inter-symbol interference (ISI).
After digital-to-analog conversion, the transmitted baseband analog temporal-domain signal of BS $m$ at time instant $t$ within TTS $n$ is written as
\begin{equation}  \label{baseband}
\mathbf{x}_m^n\left( t \right) = \mathop \sum \limits_{l = 0}^{L - 1} \mathop \sum \limits_{k = 0}^{K - 1} \mathbf{x}_{m,k}^n\left[ l \right]{e^{j2\pi k \Delta f\left( {t - l{T_{s}}} \right)}}{r}\left( {\frac{{t - l{T_{s}} + {T_{c}}}}{T_{s}}} \right),
\end{equation}
where ${r}\left( t/T_s \right)$ is a rectangular function that equals 1 when $0 \le t \le {T_{{s}}}$, and 0 otherwise.
Subsequently, the baseband signal in (\ref{baseband}) is up-converted to carrier frequency $f_m$ and transmitted through the antenna array.
\subsection{Sensing Model}
The transmitted OFDM signals are received by the BSs after being reflected by the target.
Since the BSs operate at non-overlapping frequency bands, the transmit signals from other BSs can be eliminated by bandpass filters operated at each BS. It is assumed that the BSs are capable of sensing the non-line-of-sight (NLoS) paths introduced by reflection/scattering of stationary objects in the propagation environment, such that the BSs can mitigate target-free clutters introduced by these stationary objects.
Then, after down-converting, the baseband echo signal received by BS $m$ at time instant $t$ within TTS $n$ (after clutter elimination) is expressed as
\begin{equation} \label{receive_sense}
\mathbf{y}_m^n\left( t \right) = \alpha _m^n{\mathbf{a}_r}\left( {\theta _m^n} \right)\mathbf{a}_t^H\left( {\theta _m^n} \right) \mathbf{x}_{m}^n\left( {t - \tau _m^n} \right){e^{j2\pi \mu _m^nt}} + {\mathbf{z}_m}\left( t \right),
\end{equation}
where $\alpha_m^n = \sqrt {{N_t}{N_r}} \delta_m^n{e^{- j2\pi f_m\tau_m^n}}$, with $\delta_m^n$, $\tau_m^n$, and $\mu_m^n$ denoting the reflection coefficient, the time delay, and the Doppler frequency shift between the target and BS $m$ at TTS $n$, respectively.
Specifically, the reflection coefficient is calculated by $\delta_m^n=\beta^n \sqrt{\varepsilon_m^n}$, incorporating the effects of the complex radar cross section (RCS) $\beta^n$ and round-trip path loss ${\varepsilon_m^n}$.
$\theta_m^n$ is the angle between the target and BS $m$ at TTS $n$.
Moreover, ${\mathbf{a}_t}\left( \theta  \right) = \sqrt {\frac{1}{{{N_t}}}} {\left[ {1,{e^{ - j\pi \cos \theta }}, \cdots ,{e^{ - j\left( {{N_t} - 1} \right)\pi \cos \theta }}} \right]^T}$ and ${\mathbf{a}_r}\left( \theta  \right) = \sqrt {\frac{1}{{{N_r}}}} {\left[ {1,{e^{ - j\pi \cos \theta }}, \cdots ,{e^{ - j\left( {{N_r} - 1} \right)\pi \cos \theta }}} \right]^T}$ denote the transmit and receive steering vectors of the antenna array at angle $\theta$. Besides, ${\mathbf{z}_m}\left( t \right) \sim {\cal C}{\cal N}\left( {\mathbf{0},\sigma _m^2{\mathbf{I}_{{N_r}}}} \right)$ is the complex additive white Gaussian noise (AWGN) at BS $m$.

Then, we perform analog-to-digital conversion, CP removal, and the $K$-point discrete Fourier transform (DFT) on the received baseband analog temporal-domain signal in (\ref{receive_sense}), which is converted into the baseband signal in the frequency domain.
Specifically, the frequency-domain signal received by BS $m$ at subcarrier $k$, OFDM symbol $l$, and TTS $n$ is represented as
\begin{align}  \label{sense_freq}
& \mathbf{y}_{m,k}^n\left[ l \right] = \alpha _m^n{\mathbf{a}_r}\left( {\theta_m^n} \right)\mathbf{a}_t^H\left( {\theta _m^n} \right)\mathbf{x}_{m,k}^n\left[ l \right]{e^{ - j2\pi k \Delta f\tau _m^n}}{e^{j2\pi \mu _m^nl{T_{s}}}} \nonumber \\
& \quad\quad\quad\quad + {\mathbf{z}_{m,k}}\left[ l \right].
\end{align}
We will perform target parameters estimation based on $\mathbf{y}_{m,k}^n\left[ l \right]$ in (\ref{sense_freq}) in Section III-B.
\subsection{Communication Model}
The OFDM signals transmitted by the BSs are received by the mobile device after propagating through the downlink communication channel.
We assume that the data information required by the device is only available at one service BS, which is denoted by $u \in {\cal M}$ without loss of generality. In practice, BS $u$ can be determined by the ISAC network based on the channel conditions between the BSs and the mobile device. We suppose that BS $u$ is pre-determined and leave the device-BS association issue in future work.
As different BSs operate at different frequencies, the received baseband frequency-domain signal at the mobile device\footnote{The communication channel is modeled to capture the characteristics of the high-frequency band. Due to the sparsity of the high-frequency channel, the strong LoS path may dominate other NLoS paths \cite{niu2015survey}. Besides, when the aerial platform (e.g., a UAV) serves as the mobile device operating at a higher altitude above the ground, the lack of surrounding scatterers further emphasizes the LoS dominance. Therefore, within the scope of this study, the LoS channel model is adopted to characterize the communication links.} from BS $u \in {\cal M}$ at subcarrier $k$, OFDM symbol $l$, and TTS $n$ can be expressed as
\begin{equation} \label{com}
{y}_{u,k}^n\left[ l \right] = \bar \alpha_{u}^n \mathbf{a}_t^H\left( {\theta_u^n} \right) \mathbf{x}_{u,k}^n\left[ l \right] {e^{- j2\pi k\Delta f\bar \tau _u^n}}{e^{j2\pi \bar \mu_u^n l{T_s}}} + {z_{u,k}}\left[ l \right],
\end{equation}
where $\bar \alpha_{u}^n = \sqrt {{N_t}} \sqrt {\bar \varepsilon_u^n} {e^{-j2\pi {f_u}\bar \tau_u^n}}$, ${\bar \varepsilon_u^n}$, $\bar \tau_u^n$, and $\bar \mu_u^n$ represent the path-loss coefficient, the time delay, and the Doppler frequency shift between the device and BS $u$, respectively.
Besides, ${z_{u,k}}\left[ l \right] \sim {\cal C}{\cal N}\left( {{0},\sigma_c^2} \right)$ is the AWGN at the device.

\section{Cooperative Sensing-Assisted Target Tracking}
In this section, we propose a cooperative sensing-assisted target tracking scheme to precisely track the state of the mobile device/target.
In specific, we first derive the state evolution model of the target and formulate the measurement model, and then propose the EKF-based tracking scheme.
\subsection{State Evolution Model}
The purpose of target tracking is to precisely acquire the variations of target motion parameters, which can be measured by the state vector of the target. Specifically, let ${\mathbf{d}_n} = {\left[ {\mathbf{p}_n^T, \mathbf{v}_n^T} \right]^T} = {\left[ {p_x^n,p_y^n,v_x^n,v_y^n} \right]^T}$ denote the state of the target at TTS $n$, where $\mathbf{v}_n^T = {\left[ {v_x^n,v_y^n} \right]^T}$ denotes its velocity at TTS $n$.
Then, a nearly constant velocity model \cite{stone2013bayesian} is adopted to characterize the motion of the target.
In specific, the position and velocity of the target at TTS $n$ are given by
\begin{equation} \label{state}
\begin{array}{l}
{\mathbf{p}_n} = {\mathbf{p}_{n - 1}} + {\mathbf{v}_{n - 1}}\Delta T + {\mathbf{u}_{p,n}}, \\
{\mathbf{v}_n} = {\mathbf{v}_{n - 1}} + {\mathbf{u}_{v,n}},
\end{array}
\end{equation}
where ${\mathbf{u}_{p,n}} = {[ {{u_{p_x^{n}}},{u_{p_y^{n}}}} ]^T}$ and ${\mathbf{u}_{v,n}} = {[ {{u_{v_x^{n}}},{u_{v_y^{n}}}} ]^T}$ with ${u_{p_x^{n}}}$, ${u_{p_y^{n}}}$, ${{u}_{v,n}}$, and ${u_{v_x^{n}}}$ denoting the AWGNs with zero mean and variances $\sigma _{{p_x}}^2$, $\sigma _{{p_y}}^2$, $\sigma _{{v_x}}^2$, and $\sigma _{{v_y}}^2$, respectively.
Then, based on (\ref{state}), the state evolution model is stacked as
\begin{equation}  \label{state_evolution}
{\mathbf{d}_n} = \mathbf{F}{\mathbf{d}_{n-1}} + {\mathbf{u}_{n}},
\end{equation}
where
$\mathbf{F} = {\mathbf{I}_{2 \times 2}} \otimes \left[ {\begin{array}{*{20}{c}} 1&{\Delta T}\\ 0&1 \end{array}} \right]$ denotes the transition matrix and ${\mathbf{u}_{n}} = {\left[ {\mathbf{u}_{p,n}^T,\mathbf{u}_{v,n}^T} \right]^T}$ is the AWGN with zero mean and covariance matrix ${\mathbf{Q}_u} = \textup{diag}\left( {\sigma _{{p_x}}^2,\sigma _{{p_y}}^2,\sigma _{{v_x}}^2,\sigma _{{v_y}}^2} \right)$.

\subsection{Measurement Model}
Based on the received frequency-domain signal by each BS in (\ref{sense_freq}), we aim to obtain the position-velocity information by estimating the time delay, angle, and Doppler frequency shift of the moving target at each TTS.
Then, the measurement model for the target position and velocity is obtained by establishing their relationship with the estimated parameters.
Towards this end, we propose a signal processing approach based on the 2D-DFT for the received signals at each antenna.
Specifically, the frequency-domain signal received by antenna $\iota$ of BS $m$ at subcarrier $k$, OFDM symbol $l$, and TTS $n$ is expressed as ${[ {\mathbf{y}_{m,k}^n\left[ l \right]} ]_{{\iota}}} = \alpha_m^n {e^{-j{\iota}\pi \cos \theta_m^n}} \mathbf{a}_t^H\left( {\theta_m^n} \right) \mathbf{x}_{m,k}^n\left[ l \right] {e^{-j2\pi k \Delta f\tau _m^n}} {e^{j2\pi \mu _m^nl{T_s}}}  + {\left[ {{\mathbf{z}_{m,k}}\left[ l \right]} \right]_{{\iota}}}$, $\iota\in \{0,...,N_r-1\}$.
Since the communication symbol $s_{m,k}^n\left[ l \right]$ is random and the digital beamforming $\mathbf{w}_{m,k}^n $ varies over subcarriers, the transmit signal $\mathbf{x}_{m,k}^n\left[ l \right]$ needs to be removed to eliminate its impact on the time delay and Doppler frequency shift analysis.
However, due to the multi-antenna transmission architecture of the MIMO-OFDM ISAC system, the transmit signal $\mathbf{x}_{m,k}^n\left[ l \right]$ and the steering vector $\mathbf{a}_t^H\left( {\theta _m^n} \right)$ are intricately coupled.
Note that $\mathbf{x}_{m,k}^n\left[ l \right]$ is known at BS $m$, while ${\theta_m^n}$ is an unknown parameter to be estimated.
Hence, it is imperative to initially estimate ${\theta _m^n}$. To tackle this issue, we first apply the $N_r$-point DFT on the signals of all receiving antennas along the spatial dimension, and then use peak detection to estimate ${\theta_m^n}$.
Then, the element-wise complex division between the received signal ${[ {\mathbf{y}_{m,k}^n\left[ l \right]} ]_{{\iota}}}$ and the signal-related component $\mathbf{a}_t^H\left( {\theta_m^n} \right) \mathbf{x}_{m,k}^n\left[ l \right]$ is adopted, thus obtaining
\begin{align} \label{element-wise}
&{\left[ {\mathbf{\bar y}_{m,k}^n\left[ l \right]} \right]_{{\iota}}} = {\left[ {\mathbf{y}_{m,k}^n\left[ l \right]} \right]_{{\iota}}} / \mathbf{a}_t^H\left( {\theta_m^n} \right) \mathbf{x}_{m,k}^n\left[ l \right]  \nonumber \\
&= \alpha_m^n {e^{-j{\iota}\pi \cos \theta_m^n}} {e^{-j2\pi k \Delta f\tau _m^n}}{e^{j2\pi \mu _m^nl{T_s}}} + {\left[ {{\mathbf{\bar z}_{m,k}}\left[ l \right]} \right]_{{\iota}}},
\end{align}
where ${\left[ {{{\mathbf{\bar z}}_{m,k}}\left[ l \right]} \right]_{{\iota}}} = {\left[ {{\mathbf{z}_{m,k}}\left[ l \right]} \right]_{{\iota}}} / \mathbf{a}_t^H\left( {\theta_m^n} \right) \mathbf{x}_{m,k}^n\left[ l \right]$.
For ease of representation, (\ref{element-wise}) is stacked into a compact matrix by aggregating $K$ subcarriers and $L$ symbols as
${[ {\mathbf{\bar Y}_m^n} ]_{{\iota}}} = \alpha_m^n {e^{-j{\iota}\pi \cos \theta_m^n}} {{\mathbf{\tau}}_{\tau _m^n}} {\mathbf{\tau}}_{\mu _m^n}^T + {[ {{{\mathbf{\bar Z}}_m}} ]_{{\iota}}}$,
where ${{\mathbf{\tau}}_{\tau_m^n}} = {\left[ {1,{e^{ - j2\pi \Delta f\tau _m^n}}, \cdots ,{e^{ - j2\pi \left( {K - 1} \right)\Delta f\tau _m^n}}} \right]^T}$ and ${{\mathbf{\tau}}_{\mu _m^n}} = {\left[ {1,{e^{j2\pi \mu _m^n{T_{s}}}}, \cdots ,{e^{j2\pi \mu _m^n\left( {L - 1} \right){T_{s}}}}} \right]^T}$.
It is worth pointing out that $[ {\mathbf{\bar Y}_m^n} ]_{{\iota}}$ and ${[ {{{\mathbf{\bar Z}}_m}} ]_{{\iota}}}$ are stacked such that their $\left( {k,l} \right)$-th elements are ${[ {\mathbf{\bar y}_{m,k}^n\left[ l \right]} ]_{{\iota}}}$ and ${\left[ {{\mathbf{\bar z}_{m,k}}\left[ l \right]} \right]_{{\iota}}}$, respectively.
Then, the distance and velocity can be estimated by performing the 2D-DFT on ${\left[ {\mathbf{\bar Y}_m^n} \right]_{{\iota}}}$.
Specifically, by performing the $K$-point IDFT on each column and the $L$-point DFT on each row of ${\left[ {\mathbf{\bar Y}_m^n} \right]_{{\iota}}}$, the estimations of the time delay and Doppler frequency shift, denoted as $\hat \tau _m^n$ and $\hat \mu _m^n$, respectively, can be acquired through peak detection.

Next, we obtain the measurement models of the target position $\mathbf{p}_n$ and velocity $\mathbf{v}_n$ based on their relationship with the estimations $\hat \tau _m^n$ and $\hat \mu _m^n$, which are expressed as
\begin{align}
& \hat \tau _m^n = \frac{{2\left\| {{\mathbf{p}_n} - {\mathbf{b}_m}} \right\|}}{c} + {z_{\tau_m^n}}, \label{tau}\\
& \hat \mu _m^n = \frac{{2f_m \mathbf{v}_n^T \left( {{\mathbf{p}_n} - {\mathbf{b}_m}} \right)}}{{c\left\| {{\mathbf{p}_n} - {\mathbf{b}_m}} \right\|}} + {z_{{\mu_m^n}}},  \label{mu}
\end{align}
where $c$ denotes the speed of light, ${z_{{\tau_m^n}}}$ and  ${z_{{\mu_m^n}}}$ are the AWGNs with zero mean and variance $\sigma_{{\tau_m}}^2$ and $\sigma_{{\mu _m}}^2$, respectively.
Then, based on the estimated time delay $\hat \tau _m^n$ and Doppler frequency shift $\hat \mu _m^n$, matched filtering is applied to the received signal $\mathbf{y}_{m,k}^n\left[ l \right]$ in (\ref{sense_freq}) to obtain a measurement model of the target position ${\mathbf{p}_n}$ as\footnote{Due to the geometric relationship between the target position ${\mathbf{p}_n}$ and angle ${\theta _m^n}$, i.e., ${\theta_m^n} = \arccos \frac{{{ p_x^{n}} - b_x^m}}{{\left\| {{\mathbf{p}_{n}} - {\mathbf{b}_m}} \right\|}}$, (\ref{y_mn}) can be regarded as a measurement model of the target position ${\mathbf{p}_n}$.}
\begin{equation}  \label{y_mn}
\mathbf{\tilde y}_{m,k}^n\left[ l \right] = \alpha_m^n \omega_{m,k,l}^n {\mathbf{a}_r}\left( {\theta _m^n} \right)\mathbf{a}_t^H\left( {\theta _m^n} \right) \mathbf{w}_{m,k}^n + \mathbf{\tilde z}_{m,k}^n\left[ l \right],
\end{equation}
where $\omega_{m,k,l}^n$ denotes the matched-filtering gain and $\mathbf{\tilde z}_{m,k}^n\left[ l \right]$ is the AWGN with zero mean and covariance matrix $\sigma_{{\theta_m}}^2 \mathbf{I}_{{N_r}} $.
For ease of illustration, the measurement vector and measurement noise of ${\mathbf{p}_n}$ are stacked as
$\mathbf{\tilde y}_m^n = {[ {{( {\mathbf{\tilde y}_{m,0}^n\left[ 0 \right]} )^T}, \cdots, {{( {\mathbf{\tilde y}_{m,K - 1}^n\left[ {L - 1} \right]} )}^T}} ]^T} \in \mathbb{C}{^{LK{N_r} \times 1}}$ and $\mathbf{z}_{\theta _m^n} = {[ {{( {\mathbf{\tilde z}_{m,0}^n\left[ 0 \right]} )^T}, \cdots, {{( {\mathbf{\tilde z}_{m,K - 1}^n\left[ {L - 1} \right]} )}^T}} ]^T} \in \mathbb{C}{^{LK{N_r} \times 1}}$, respectively.
Let the measurement vector of BS $m$ be denoted as $\mathbf{r}_m^n = {[ {\hat \tau _m^n,\hat \mu _m^n,{{\left( {\mathbf{\tilde y}_m^n} \right)}^T}} ]^T}$.
Then, based on (\ref{tau})-(\ref{y_mn}), the measurement model of the target state is stacked as
\begin{equation} \label{measurement_model}
\mathbf{r}_m^n = {\mathbf{g}_m}\left( {{\mathbf{d}_n}} \right) + \mathbf{z}_m^n,
\end{equation}
where ${\mathbf{g}_m}\left( \cdot \right)$ is define according to (\ref{tau})-(\ref{y_mn}) and $\mathbf{z}_m^n = {[ {{z_{\tau _m^n}},{z_{\mu _m^n}},\mathbf{z}_{\theta _m^n}^T} ]^T}$ is the measurement noise with zero mean and covariance matrix ${\mathbf{Q}_{{z_m}}} = \textup{diag} \left( {\sigma _{{\tau _m}}^2,\sigma _{{\mu _m}}^2,\sigma _{{\theta _m}}^2 \mathbf{1}_{{LKN_r}}^T} \right)$.

\subsection{Extended Kalman Filtering (EKF) for Target Prediction}
In this subsection, according to the derived state evolution model in (\ref{state_evolution}) and measurement model in (\ref{measurement_model}), a Kalman filtering method is proposed to fuse the measurement results from multiple BSs, thereby predicting the target state for predictive beamforming design next.
Since the measurement model in (\ref{measurement_model}) exhibits nonlinearity, the EKF method is utilized for estimating and predicting the target state by linearizing the nonlinear model in (\ref{measurement_model}).
For the convenience of representation, based on (\ref{y_mn}), we define
\begin{align}
{\mathbf{o}_{m,k,l}}\left( {{\theta_m}} \right) = {\alpha_m} \omega_{m,k,l} {\mathbf{a}_r}\left( {{\theta _m}} \right)\mathbf{a}_t^H\left( {{\theta _m}} \right) {\mathbf{w}_{m,k}}.
\end{align}
Then, to linearize the measurement model in (\ref{measurement_model}), the Jacobian matrix for $\mathbf{g}_m$ is derived as
\begin{align}   \label{G_m}
{\mathbf{G}_m} =
\frac{{\partial {\mathbf{g}_m}}}{\mathbf{\partial d}} = \left[ {\begin{array}{*{20}{c}}
{\frac{{\partial {\tau_m}}}{{\partial {p_x}}}}&{\frac{{\partial {\tau_m}}}{{\partial {p_y}}}}&0&0\\
{\frac{{\partial {\mu_m}}}{{\partial {p_x}}}}&{\frac{{\partial {\mu_m}}}{{\partial {p_y}}}}&{\frac{{\partial {\mu_m}}}{{\partial {v_x}}}}&{\frac{{\partial {\mu_m}}}{{\partial {v_y}}}}\\
{\frac{{\partial {\mathbf{o}_m}}}{{\partial {p_x}}}}&{\frac{{\partial {\mathbf{o}_m}}}{{\partial {p_y}}}}&0&0
\end{array}} \right],
\end{align}
where ${\mathbf{o}_m}={\left[ {{{\left( {\mathbf{o}_{m,0,0}^T} \right)}}, \cdots, {{\left( {\mathbf{o}_{m,K-1,L-1}^T} \right)}}} \right]^T} \in \mathbb{C}{^{LK{N_r} \times 1}}$.
In (\ref{G_m}), we have
$\frac{{\partial {\tau_m}}}{{\partial {{p}_x}}} = \frac{{2 \ell_{x,m} }}{{c\left\| {\mathbf{p} - {\mathbf{b}_m}} \right\|}}$,
$\frac{{\partial {\tau_m}}}{{\partial {p_y}}} = \frac{{2 \ell_{y,m} }}{{c\left\| {\mathbf{p} - {\mathbf{b}_m}} \right\|}}$,
$\frac{{\partial {\mu_m}}}{{\partial {p_x}}} = \frac{{2f_m \ell_{y,m} \psi_{x,y,m}}}{{c{{\left\| {\mathbf{p} - {\mathbf{b}_m}} \right\|}^3}}}$, $\frac{{\partial {\mu_m}}}{{\partial {p_y}}} = -\frac{{2{f_m} \ell_{x,m} \psi _{x,y,m}}}{{c{{\left\| {\mathbf{p} - {\mathbf{b}_m}} \right\|}^3}}}$, $\frac{{\partial {\mu_m}}}{{\partial {v_x}}} = \frac{{2{f_m} \ell_{x,m} }}{{c\left\| {\mathbf{p} - {\mathbf{b}_m}} \right\|}}$,
$\frac{{\partial {\mu_m}}}{{\partial {v_y}}} = \frac{{2{f_m} \ell_{y,m} }}{{c\left\| {\mathbf{p} - {\mathbf{b}_m}} \right\|}}$,
$\frac{{\partial {\mathbf{o}_{m}}}}{{\partial {p_x}}} = {\left[ {{{ {\frac{{\partial {\mathbf{o}_{m,0,0}^T}}}{{\partial {p_x}}}} }}, \cdots, {{ {\frac{{\partial {\mathbf{o}_{m,K-1,L-1}^T}}}{{\partial {p_x}}}} }}} \right]^T}$,
and $\frac{{\partial {\mathbf{o}_{m}}}}{{\partial {p_y}}}=$ ${\left[ {{{{\frac{{\partial {\mathbf{o}_{m,0,0}^T}}}{{\partial {p_y}}}} }}, \cdots, {{ {\frac{{\partial {\mathbf{o}_{m,K-1,L-1}^T}}}{{\partial {p_y}}}} }}} \right]^T}$,
where $\ell_{x,m} = p_x - b_x^m$, $\ell_{y,m} = p_y - b_y^m$, $\psi_{x,y,m} = v_x \ell_{y,m} - v_y \ell_{x,m}$, $\frac{{\partial {\mathbf{o}_{m,k,l}}}}{{\partial {p_x}}} = \mathbf{ {\cal A}}_{m,k,l}^y{\mathbf{w}_{m,k}}$, and
$\frac{{\partial {\mathbf{o}_{m,k,l}}}}{{\partial {p_y}}} = \mathbf{ {\cal A}}_{m,k,l}^x{\mathbf{w}_{m,k}}$.
Specifically, ${\cal A}_{m,k,l}^x = {\alpha_m}{\omega_{m,k,l}}\frac{{{p_x} - b_x^m}}{{{{\left\| {\mathbf{p} - {\mathbf{b}_m}} \right\|}^2}}}{\dot {\mathbf{A}}}\left( {{\theta_m}} \right)$ and ${\cal A}_{m,k,l}^y = {\alpha_m}{\omega_{m,k,l}}\frac{{b_y^m - {p_y}}}{{{{\left\| {\mathbf{p} - {\mathbf{b}_m}} \right\|}^2}}}{\dot {\mathbf{A}}}\left( {{\theta _m}} \right)$, where $\dot {\mathbf{A}}\left( {{\theta_m}} \right) = {\dot {\mathbf{a}}_r}\left( {{\theta_m}} \right)\mathbf{a}_t^H\left( {{\theta _m}} \right) + {\mathbf{a}_r}\left( {{\theta_m}} \right)\dot {\mathbf{a}}_t^H\left( {{\theta_m}} \right)$ is the derivative of ${\mathbf{A}}\left( {{\theta_m}} \right)$ with respect to ${\theta_m}$,
with ${{\dot {\mathbf{a}}}_t}\left( {{\theta_m}} \right) = {{\bar {\mathbf{a}}}_t}\left( {{\theta_m}} \right) \odot {\mathbf{a}_t}\left( {{\theta _m}} \right)$ and ${{\dot {\mathbf{a}}}_r}\left( {{\theta _m}} \right) = {{\bar {\mathbf{a}}}_r}\left( {{\theta _m}} \right) \odot {\mathbf{a}_r}\left( {{\theta _m}} \right)$ denoting the derivatives of ${{{\mathbf{a}}}_t}\left( {{\theta _m}} \right)$ and ${{\mathbf{a}}_r}\left( {{\theta _m}} \right)$ with respect to ${\theta_m}$, respectively.
Moreover, ${\bar {\mathbf{a}}_t}\left( {{\theta_m}} \right) = {[0,j\pi \sin {\theta _m}, \cdots ,j\left( {{N_t} - 1} \right)\pi \sin {\theta _m}]^T}$ and
${\bar {\mathbf{a}}_r}\left( {{\theta _m}} \right) = {[0,j\pi \sin {\theta_m}, \cdots ,j\left( {{N_r} - 1} \right)\pi \sin {\theta _m}]^T}$.

After developing the state evolution model and linearizing the measurement model, the procedure of the EKF method \cite{kay1993fundamentals} is presented as follows:
\subsubsection{State Prediction} ${\mathbf{\hat d}_{n\left| {n-1} \right.}} = \mathbf{F} {\mathbf{\hat d}_{n-1}}$.
\subsubsection{Covariance Matrix Prediction} ${\mathbf{M}_{n\left| {n - 1} \right.}} = \mathbf{F}{\mathbf{M}_{n - 1}}{\mathbf{F}^H} + {\mathbf{Q}_u}$,
where ${\mathbf{M}_{n\left| {n - 1} \right.}}$ and ${\mathbf{M}_{n - 1}}$ denote the predictive covariance matrix and the updated covariance matrix for target state estimation at TTS $n-1$, respectively.
\subsubsection{Kalman Gain Calculation} ${\mathbf{K}_n} = {\mathbf{M}_{n\left| {n-1} \right.}}\mathbf{G}_n^H \left( {\mathbf{G}_n}{\mathbf{M}_{n\left| {n-1} \right.}}\mathbf{G}_n^H + {\mathbf{Q}_z} \right)^{-1}$,
where ${\mathbf{G}_n}=$ ${[ {{{\left( {\mathbf{G}_1^n} \right)}^H}, \cdots ,{{\left( {\mathbf{G}_M^n} \right)}^H}} ]^H}$ with $\mathbf{G}_m^n = \frac{{\partial {\mathbf{g}_m}}}{{\partial \mathbf{d}}}\left| {_{\mathbf{d} = {\mathbf{\hat{d}}_{n\left| {n - 1} \right.}}}} \right.$, and ${\mathbf{Q}_z} = {\rm{blkdiag}}\left( {{\mathbf{Q}_{{z_1}}}, \cdots ,{\mathbf{Q}_{{z_M}}}} \right)$.
\subsubsection{State Estimation Update} ${\mathbf{\hat d}_n} = {\mathbf{\hat d}_{n\left| {n - 1} \right.}} + {\mathbf{K}_n}\left( {{\mathbf{r}_n} - {\mathbf{r}_{n\left| {n - 1} \right.}}} \right)$,
where ${\mathbf{r}_n} = {[ {{{\left( {\mathbf{r}_1^n} \right)}^H}, \cdots ,{{\left( {\mathbf{r}_M^n} \right)}^H}} ]^H}$ and ${\mathbf{r}_{n\left| {n-1} \right.}} = {[ {\mathbf{g}_1^H( {{{\hat {\mathbf{d}}}_{n\left| {n-1} \right.}}} ), \cdots , \mathbf{g}_M^H ( {{{\hat {\mathbf{d}}}_{n\left| {n-1} \right.}}} )} ]^H}$.
In this step, the individual measurement results from different BSs are fused for joint estimation.
\subsubsection{Covariance Matrix Update} ${\mathbf{M}_n} = \left( {\mathbf{I} - {\mathbf{K}_n}{\mathbf{G}_n}} \right){\mathbf{M}_{n\left| {n-1} \right.}}$,
where ${\mathbf{M}_{n}}$ denotes the updated covariance matrix at TTS $n$.

Before employing the EKF method to estimate and predict the target, obtaining its initial state is essential, which can be achieved through conventional beam training.
Based on the predicted target state ${\mathbf{\hat d}_{n\left| {n-1} \right.}}$ obtained by the EKF method, the predictive beamforming can be designed to facilitate high-quality communication and high-accuracy estimation for the moving target, as shown in the next section.

\section{Predictive Beamforming Design}
This section presents a predictive beamforming design to maximize communication performance of the device while ensuring sensing performance.
First, we derive the communication and sensing metrics in the subsequent TTS, and then formulate the optimization problem for predictive beamforming. Accordingly, an SDR-based algorithm and a penalty-based algorithm are developed to solve the formulated problem, respectively.

\subsection{Performance Metric Derivation}

\subsubsection{Achievable Rate for Communication}
Based on the signal model in (\ref{com}), the achievable rate of the device at subcarrier $k$ and TTS $n$ is
\begin{equation}  \label{rate}
R_{u,k}^n = \textup{log}{_2}\left( {1 + {{\left| {{\left( {\mathbf{h}_{u,k}^n} \right)^H}\mathbf{w}_{u,k}^n} \right|}^2}/\sigma_{c}^2} \right),
\end{equation}
where ${\mathbf{h}_{u,k}^n} = \bar \alpha_{u}^n {e^{- j2\pi k\Delta f\bar \tau_u^n}} \mathbf{a}_t\left( {\theta_u^n} \right) \in \mathbb{C}^{N_t \times 1}$ with $\bar \alpha_{u}^n = \sqrt {{N_t}} \sqrt {\bar \varepsilon_u^n} {e^{-j2\pi {f_u}\bar \tau_u^n}}$ is the equivalent channel vector between BS $u$ and the device at subcarrier $k$,
which is constructed based on the obtained target state prediction ${\mathbf{\hat d}_{n\left| {n-1} \right.}} = {\left[ {\mathbf{\hat p}_{n\left| {n-1} \right.}^T, \mathbf{\hat v}_{n\left| {n-1} \right.}^T} \right]^T} = {\left[ {\hat p_x^{n\left| {n-1} \right.}, \hat p_y^{n\left| {n-1} \right.}, \hat v_x^{n\left| {n-1} \right.}, \hat v_y^{n\left| {n-1} \right.}} \right]^T}$.
Specifically, the path-loss coefficient ${\bar \varepsilon_u^n}$ is determined by the distance ${{\left\| {{\mathbf{\hat p}_{n\left| {n-1} \right.}} - {\mathbf{b}_u}} \right\|}}$ between the device and BS $u$ at TTS $n$.
Furthermore, the time delay and angle between the device and BS $u$ at TTS $n$ are calculated as $\bar \tau_u^n = \frac{{\left\| {{\mathbf{\hat p}_{n\left| {n-1} \right.}} - {\mathbf{b}_u}} \right\|}}{c}$ and ${\theta_u^n} = \arccos \frac{{{\hat p_x^{n\left| {n-1} \right.}} - b_x^u}}{{\left\| {{\mathbf{\hat p}_{n\left| {n-1} \right.}} - {\mathbf{b}_u}} \right\|}}$, $u \in {\cal M}$, respectively.

\subsubsection{PC-CRLB for Target Position Estimation}
The CRLB is commonly utilized to quantify the estimation performance, serving as a lower bound on the variance of any unbiased estimators \cite{kay1993fundamentals}.
Unlike the conventional CRLB solely depending on measurement results, target tracking performance requires the joint consideration of both the measurement model in (\ref{measurement_model}) and the state model in (\ref{state_evolution}).
Therefore, the PC-CRLB is proposed to average over the joint density of the state $\mathbf{d}_n$ and the measurement result $\mathbf{r}_n$ conditioned on the actual past measurements ${\mathbf{r}_{1:n - 1}}$ \cite{bell2015cognitive}.
Specifically, the mean squared error (MSE) matrix of an estimate ${\mathbf{\bar {d}}}_n$ for the target state $\mathbf{d}_n$ based on the observation ${\mathbf{r}_{n}}$ is lower bounded by the PC-CRLB matrix ${\mathbf{C}_n}\left( {\mathbf{d}_n} \right)$ as
\begin{align}
&{\mathbb{E}_{{\mathbf{d}_n},{\mathbf{r}_n}\left| {{\mathbf{r}_{1:n - 1}}} \right.}}\left\{ {\left( {{\mathbf{\bar d}_n} - {\mathbf{d}_n}} \right){{\left( {{\mathbf{\bar d}_n} - {\mathbf{d}_n}} \right)}^T}} \right\} \nonumber \\
& \underset{\scriptscriptstyle}{\succeq} {\mathbf{J}^{ - 1}}\left( {{\mathbf{d}_n}\left| {{\mathbf{r}_{1:n - 1}}} \right.} \right) \buildrel \Delta \over = {\mathbf{C}_n}\left( {\mathbf{d}_n} \right),
\end{align}
where ${\mathbf{J}}\left( {{\mathbf{d}_n}\left| {{\mathbf{r}_{1:n - 1}}} \right.} \right)$ is the predicted conditional Fisher information matrix (PC-FIM).
Then, according to \cite{bell2015cognitive}, the PC-FIM is expanded as
\begin{align}  \label{OP}
\mathbf{J}\left( {{\mathbf{d}_n}\left| {{\mathbf{r}_{1:n - 1}}} \right.} \right)
& \buildrel \Delta \over = {\mathbf{J}_P}\left( {{\mathbf{d}_n}\left| {{\mathbf{r}_{1:n - 1}}} \right.} \right) + {\mathbf{J}_O}\left( {{\mathbf{d}_n}\left| {{\mathbf{r}_{1:n - 1}}} \right.} \right),
\end{align}
where ${\mathbf{J}_P}\left( {{\mathbf{d}_n}\left| {{\mathbf{r}_{1:n - 1}}} \right.} \right) = {-\mathbb{E}{_{{\mathbf{d}_n}\left| {{\mathbf{r}_{1:n - 1}}} \right.}}\left( {\frac{{{\partial ^2}\ln \pi \left( {{\mathbf{d}_n}\left| {{\mathbf{r}_{1:n - 1}}} \right.} \right)}}{{\partial {\mathbf{d}_n}\partial \mathbf{d}_n^T}}} \right)}$ is the prior FIM with $\pi \left( {{\mathbf{d}_n}\left| {{\mathbf{r}_{1:n - 1}}} \right.} \right)$ denoting the priori probability density function (PDF) of ${\mathbf{d}_n}$ and ${\mathbf{J}_O}\left( {{\mathbf{d}_n}\left| {{\mathbf{r}_{1:n - 1}}} \right.} \right) = { - \mathbb{E} {_{{\mathbf{d}_n},{\mathbf{r}_n}\left| {{\mathbf{r}_{1:n - 1}}} \right.}}\left( {\frac{{{\partial ^2}\ln \hat f\left( {{\mathbf{r}_n}\left| {{\mathbf{d}_n};{\mathbf{r}_{1:n - 1}}} \right.} \right)}}{{\partial {\mathbf{d}_n}\partial \mathbf{d}_n^T}}} \right)}$ is the observed FIM with $\hat f\left( {{\mathbf{r}_n}\left| {{\mathbf{d}_n}; {\mathbf{r}_{1:n - 1}}} \right.} \right)$ denoting the conditional PDF of ${\mathbf{r}_n}$ on ${\mathbf{d}_n}$.
Since the measurement of each BS is independent, we obtain $\hat f\left( {{\mathbf{r}_n}\left| {{\mathbf{d}_n};{\mathbf{r}_{1:n-1}}} \right.} \right) = \mathop \Pi \limits_{m \in {\cal M}} \hat f\left( {\mathbf{r}_m^n \left| {{\mathbf{d}_n};{\mathbf{r}_{1:n-1}}} \right.} \right)$, with $\hat f\left( {\mathbf{r}_m^n \left| {{\mathbf{d}_n};{\mathbf{r}_{1:n-1}}} \right.} \right)$ denoting the conditional PDF of ${\mathbf{r}_m^n}$ on ${\mathbf{d}_n}$.
Based on (\ref{OP}), the PC-FIM incorporates both the state evolution model and the measurement model, conditioned on past measurements. Thus, compared to the classical CRLB, the PC-CRLB offers a more accurate lower bound on the estimation error of the target state.
However, it is typically difficult to evaluate the prior FIM analytically.
For ease of analysis, the priori PDF is assumed to follow the Gaussian distribution.
Then, ${\mathbf{J}_P}\left( {{\mathbf{d}_n}\left| {{\mathbf{r}_{1:n - 1}}} \right.} \right)$ can be approximated as
the inverse of the predictive covariance matrix \cite{bell2015cognitive} as
\begin{equation} \label{JP}
{\mathbf{J}_P}\left( {{\mathbf{d}_n}\left| {{\mathbf{r}_{1:n - 1}}} \right.} \right) \approx \mathbf{M}_{n\left| {n - 1} \right.}^{-1}.
\end{equation}
Based on the measurement model and considering the independence of measurement results from different BSs, the observed FIM is expressed as ${\mathbf{J}_O}\left( {{\mathbf{d}_n}\left| {{\mathbf{r}_{1:n - 1}}} \right.} \right) = \sum\limits_{m = 1}^M {\mathbb{E}_{{\mathbf{d}_n}\left| {{\mathbf{r}_{1:n - 1}}} \right.}} \left\{ {{\left( {\mathbf{G}_m^n} \right)^H}\mathbf{Q}_{{\mathbf{z}_m}}^{ - 1}\mathbf{G}_m^n} \right\}$ \cite{xie2017joint}.
Note that the observed FIM includes the expectation operator with respect to ${\mathbf{d}_n}$ conditioned on ${\mathbf{r}_{1:n - 1}}$.
To achieve real-time target tracking and under the assumption of small process noise \cite{glass2011mimo},
the observed FIM can be approximated by dropping the expectation as
\begin{align}  \label{J_O}
&{\mathbf{J}_O}\left( {{\mathbf{d}_n}\left| {{\mathbf{r}_{1:n - 1}}} \right.} \right) \approx \sum\limits_{m = 1}^M {{{\left( {\mathbf{G}_m^n} \right)}^H}\mathbf{Q}_{{\mathbf{z}_m}}^{ - 1}\mathbf{G}_m^n}.
\end{align}
Then, to facilitate the derivation of the PC-CRLB matrix for estimating the target position, the PC-FIM is partitioned into blocks as
\begin{align}
\mathbf{J} \left( {{\mathbf{d}_n}\left| {{\mathbf{r}_{1:n-1}}} \right.} \right) = \left[ {\begin{array}{*{20}{c}}
{{\mathbf{J}_{{\mathbf{p}_n}{\mathbf{p}_n}}}}&{{\mathbf{J}_{{\mathbf{p}_n}{\mathbf{v}_n}}}}\\
{\mathbf{J}_{{\mathbf{p}_n}{\mathbf{v}_n}}^T}&{{\mathbf{J}_{{\mathbf{v}_n}{\mathbf{v}_n}}}}
\end{array}} \right],
\end{align}
where the elements for the sub-matrices of $\mathbf{J} \left( {{\mathbf{d}_n}\left| {{\mathbf{r}_{1:n-1}}} \right.} \right)$ are derived in (\ref{a1})-(\ref{a2}) in the Appendix.
It is worth pointing out that each element of ${\mathbf{J}_{{\mathbf{p}_n}{\mathbf{p}_n}}}$ is a quadratic function of ${\mathbf{w}_n} = \left\{ {\mathbf{w}_1^n, \cdots,\mathbf{w}_M^n} \right\}$ with $\mathbf{w}_m^n = \left\{ {\mathbf{w}_{m,0}^n, \cdots,\mathbf{w}_{m,K-1}^n} \right\}$.
Then, the PC-CRLB matrix of the target position ${\mathbf{p}_n}$ is calculated as
\begin{align}
{\mathbf{C}_n}\left( {{\mathbf{p}_n}} \right) = {\left( {{\mathbf{J}_{{\mathbf{p}_n}{\mathbf{p}_n}}} - {\mathbf{J}_{{\mathbf{p}_n}{\mathbf{v}_n}}}\mathbf{J}_{{\mathbf{v}_n}{\mathbf{v}_n}}^{-1} \mathbf{J}_{{\mathbf{p}_n}{\mathbf{v}_n}}^T} \right)^{-1}}.
\end{align}

\subsection{Problem Formulation}
The diagonal elements of the PC-CRLB matrix stand for the minimum variance of the corresponding parameters estimated by unbiased estimators.
To evaluate sensing performance, we use the trace of the PC-CRLB matrix, which characterizes the lower bound of the sum variance for estimating different elements \cite{10217169}.
We aim to maximize the achievable rates over all subcarriers by designing the predictive beamforming, while guaranteeing the PC-CRLB requirement
and the transmit power constraints of the BSs. Specifically, the PC-CRLB-constrained rate maximization problem is
\begin{subequations} \label{problem1}
\begin{align}
&\max \limits_{{\mathbf{w}_n}} \sum\limits_{k = 0}^{K-1} {{R}_{u,k}^n}   \label{obj1}   \\
&\;\; \textup{s.t.} \;\; \mathrm{Tr}\{ {\mathbf{C}_n}\left( {{\mathbf{p}_n}} \right) \} \le \eta,  \label{constraint1}  \\
&\quad\quad \sum\limits_{k = 0}^{K-1} {\left\| {\mathbf{w}_{m,k}^n} \right\|^2} \le {P_m},\forall m \in {\cal M}, \label{pow}
\end{align}
\end{subequations}
where $\eta$ denotes the PC-CRLB threshold, ${P_m}$ denotes the transmit power budget of BS $m$, and ${{R}_{u,k}^n}$ in (\ref{obj1}) is defined in (\ref{rate}).
Notice that problem (\ref{problem1}) is non-convex, as the objective function in (\ref{obj1}) is non-concave with respect to ${\mathbf{w}_n}$ and the PC-CRLB constraint in (\ref{constraint1}) includes quadratic terms of ${\mathbf{w}_n}$.
In the following, the Schur complement is utilized to convert the PC-CRLB constraint into a more tractable form to facilitate the solution.
Then, an SDR-based algorithm and a penalty-based algorithm are proposed to tackle the reformulated problem, respectively.

Specifically, it is observed that ${\mathbf{J}_{{\mathbf{p}_n}{\mathbf{p}_n}}} - {\mathbf{J}_{{\mathbf{p}_n}{\mathbf{v}_n}}}\mathbf{J}_{{\mathbf{v}_n}{\mathbf{v}_n}}^{ - 1}\mathbf{J}_{{\mathbf{p}_n}{\mathbf{v}_n}}^T$ is
a positive semidefinite matrix. Besides, given matrix $\mathbf{A}$, $\textup{Tr}\left(\mathbf{A}^{-1}\right)$ is decreasing on the positive semidefinite matrix space. Therefore, by introducing an auxiliary matrix $\mathbf{\Omega}_n \in \mathbb{C}^{2 \times 2}$, the PC-CRLB constraint (\ref{constraint1}) is reformulated as \cite{STARS10050406}
\begin{align}
&\mathrm{Tr}\left[ {\mathbf{\Omega}_n^{-1}} \right] \le \eta ,
{\mathbf{\Omega}_n}\underset{\scriptscriptstyle}{\succeq }\mathbf{0},  \label{omega} \\
&{\mathbf{J}_{{\mathbf{p}_n}{\mathbf{p}_n}}} - {\mathbf{J}_{{\mathbf{p}_n}{\mathbf{v}_n}}}\mathbf{J}_{{\mathbf{v}_n}{\mathbf{v}_n}}^{ - 1}\mathbf{J}_{{\mathbf{p}_n}{\mathbf{v}_n}}^T \underset{\scriptscriptstyle}{\succeq } {\mathbf{\Omega}_n}.  \label{JJ}
\end{align}
According to the Schur complement \cite{zhang2006schur}, (\ref{JJ}) is converted to
\begin{align} \label{Schur}
\left[ {\begin{array}{*{20}{c}}
{{\mathbf{J}_{{\mathbf{p}_n}{\mathbf{p}_n}}} - {\mathbf{\Omega}_n}}&{{\mathbf{J}_{{\mathbf{p}_n}{\mathbf{v}_n}}}}\\
{\mathbf{J}_{{\mathbf{p}_n}{\mathbf{v}_n}}^T}&{{\mathbf{J}_{{\mathbf{v}_n}{\mathbf{v}_n}}}}
\end{array}} \right] \underset{\scriptscriptstyle}{\succeq } \mathbf{0}.
\end{align}
Note that (\ref{omega}) is convex, but (\ref{Schur}) is non-convex with respect to ${\mathbf{w}_n}$.
Next, we focus on addressing the non-convexity in (\ref{obj1}) and (\ref{Schur}).

\subsection{SDR-Based Algorithm}
We use the idea of SDR method to handle (\ref{obj1}), (\ref{pow}), and (\ref{Schur}). Towards this end, we define $\mathbf{W}_{m,k}^n = \mathbf{w}_{m,k}^n{( {\mathbf{w}_{m,k}^n} )^H}$ with $\mathbf{W}_{m,k}^n \underset{\scriptscriptstyle}{\succeq } \mathbf{0}$ and $\textup{rank}( {\mathbf{W}_{m,k}^n} ) = 1$. Then, problem (\ref{problem1}) is equivalently formulated as
\begin{subequations}  \label{SDP}
\begin{align}
&\max\limits_{\left\{ {\mathbf{W}_{m,k}^n} \right\},{\mathbf{\Omega}_n}}  \sum\limits_{k = 0}^{K - 1} {\textup{log}_2\left( {1 + \textup{Tr}\left( {\mathbf{H}_{u,k}^n\mathbf{W}_{u,k}^n} \right)/\sigma_{c}^2} \right)} \\
&\quad\quad \textup{s.t.} \quad\;\, \textup{ (\ref{omega}), (\ref{Schur})},  \\
&\quad\quad\quad\quad\;\;\, \sum\limits_{k = 0}^{K - 1} {\mathrm{Tr}\left( {\mathbf{W}_{m,k}^n} \right)}  \le {P_m},\forall m \in {\cal M},\\
&\quad\quad\quad\quad\;\;\;\, \mathbf{W}_{m,k}^n \underset{\scriptscriptstyle}{\succeq } \mathbf{0},\forall m \in {\cal M},\forall k \in {\cal K}, \\
&\quad\quad\quad\quad\;\;\;\, \textup{rank} \left( {\mathbf{W}_{m,k}^n} \right) = 1, \forall m \in {\cal M}, \forall k \in {\cal K}, \label{rank}
\end{align}
\end{subequations}
where $\mathbf{H}_{u,k}^n=\mathbf{h}_{u,k}^n{( {\mathbf{h}_{u,k}^n} )^H}$.
It is observed that problem (\ref{SDP}) is still non-convex due to the rank-one constraints in (\ref{rank}).
To address this issue, the SDR method is utilized to remove these constraints, thus obtaining a convex semi-definite programming (SDP) problem,  denoted as problem (SDR1). This problem can be optimally solved by convex optimization solvers such as CVX by using the interior-point method \cite{grant2014cvx}.
If the obtained solution $\{ {\tilde {\mathbf{W}}_{m,k}^{n}} \}$ to problem (SDR1) is rank-one, the optimal solution to problem (\ref{problem1}) is obtained by eigenvalue decomposition.
Next, we aim to reveal the tightness of the SDR method, indicating that problem (\ref{problem1}) always has an optimal solution.

{\textit{Theorem 1}}: Supposing that problem (SDR1) is feasible, there always exists an optimal solution $\{ {\tilde {\mathbf{W}}_{m,k}^{n}} \}$ satisfying $\textup{rank}({\tilde {\mathbf{W}}_{m,k}^n}) = 1$, $\forall k\in \mathcal{K}$, $\forall m\in \mathcal{M}$.

{\textit{Proof}}: Theorem 3.2 in \cite{5233822} indicates that problem (SDR1) always has an optimal solution $\{ {\tilde {\mathbf{W}}_{m,k}^{n}} \}$ with $\sum\limits_{m = 1}^M \sum\limits_{k = 0}^{K-1} {\textup{rank}{^2}( {\tilde {\mathbf{W}}_{m,k}^n} ) \le {M+K} } $. It is clear that any optimal solution satisfying $\textup{rank}( {\tilde {\mathbf{W}}_{m,k}^n}) \ge 1$, $\forall k\in \mathcal{K}$, $\forall m\in \mathcal{M}$. Hence, problem (SDR1) always has an optimal solution $\{ {\tilde {\mathbf{W}}_{m,k}^{n}} \}$ satisfying $\textup{rank}({\tilde {\mathbf{W}}_{m,k}^n}) = 1$, $\forall k\in \mathcal{K}$, $\forall m\in \mathcal{M}$.
\subsection{Penalty-Based Algorithm}
Although the SDR-based algorithm achieves a global optimum for problem (\ref{problem1}), the computational complexity of solving the SDP problem is relatively high due to the high dimension of the optimization variables.
Hence, next, we introduce an auxiliary variable and then propose a low-complexity penalty-based algorithm.
\subsubsection{Problem Reformulation}
We introduce an auxiliary variable ${\mathbf{q}_n} = {\left[ {q_n^1,q_n^2,q_n^3} \right]^T}$ and define ${\cal I}=\left\{ 1,2,3 \right\}$, thus problem (\ref{problem1}) is converted into
\begin{subequations}  \label{penal_re}
\begin{align}
&\max \limits_{{\mathbf{w}_n},{\mathbf{\Omega}_n},{\mathbf{q}_n}} \sum\limits_{k = 0}^{K - 1} R_{u,k}^n  \label{obj2}  \\
&\quad\;\, \textup{s.t.} \quad\;\, \textup{(\ref{pow})}, \textup{(\ref{omega})},  \\
&\quad\quad\quad\;\; \left[ {\begin{array}{*{20}{c}} {\left[ {\begin{array}{*{20}{c}}
{q_n^1}&{q_n^2}\\{q_n^2}&{q_n^3} \end{array}} \right] - {\mathbf{\Omega}_n}}&{{\mathbf{J}_{{\mathbf{p}_n}{\mathbf{v}_n}}}} \\
{\mathbf{J}_{{\mathbf{p}_n}{\mathbf{v}_n}}^T}&{{\mathbf{J}_{{\mathbf{v}_n}{\mathbf{v}_n}}}} \end{array}} \right] \underset{\scriptscriptstyle}{\succeq }  \mathbf{0}, \label{schur} \\
&\quad\quad\quad\quad q_n^i = f^i\left( {{\mathbf{w}_n}} \right),i \in {\cal I},  \label{penalty}
\end{align}
\end{subequations}
where \resizebox{0.4374\textwidth}{!}{$f^i\left( {{\mathbf{w}_n}} \right) = {\sum\limits_{m=1}^M \sum\limits_{k=0}^{K-1} \sum\limits_{l=0}^{L-1} {\sigma_{{\theta_m}}^{-2}\mathrm{Tr}\left( { \tilde{\cal A}_{m,k,l}^{n,i} \mathbf{w}_{m,k}^n{( {\mathbf{w}_{m,k}^n} )^H}} \right)}} + \upsilon_n^i$} with $\tilde{\cal A}_{m,k,l}^{n,i}$ and $\upsilon_n^i$ defined in the Appendix.
Then, the equality constraints in (\ref{penalty}) are added to the objective function (\ref{obj2}), thereby yielding the following problem:
\begin{subequations}  \label{add_penalty}
\begin{align}
&\min \limits_{{\mathbf{w}_n},{\mathbf{\Omega}_n},{\mathbf{q}_n}} -\sum\limits_{k = 0}^{K - 1} R_{u,k}^n + \frac{1}{{2\rho}}\sum\limits_{i = 1}^3 {{{\left| {f^i\left( {{\mathbf{w}_n}} \right) - q_n^i} \right|}^2}} \\
&\quad\;\, \textup{s.t.} \quad\;\, \textup{(\ref{pow})}, \textup{(\ref{omega})}, \textup{(\ref{schur})},
\end{align}
\end{subequations}
where $\rho>0$ is the penalty factor penalizing the violation of (\ref{penalty}).
The choice of $\rho$ has a significant impact on the objective function and the convergence behavior of the penalty-based algorithm.
Typically, $\rho$ starts with a relatively large value to establish a good initial point, and then gradually decreases until $\rho \to 0$ ($ 1/\rho \to \infty$) in the outer layer optimization. Thus, the equality constraints in (\ref{penalty}) is ensured by addressing problem (\ref{add_penalty}).
However, given $\rho$, problem (\ref{add_penalty}) is still non-convex due to multi-variable coupling.
Next, we divide problem (\ref{add_penalty}) into two sub-problems in the inner layer optimization and then solving them alternately until achieving convergence.

\subsubsection{Inner Layer Optimization}
With fixed $\mathbf{w}_n$, the sub-problem of optimizing ${\mathbf{\Omega}_n}$ and ${\mathbf{q}_n}$ is formulated as
\begin{subequations}  \label{omega_a}
\begin{align}
&\min \limits_{{\mathbf{\Omega}_n},{\mathbf{q}_n}} \frac{1}{{2\rho }}\sum\limits_{i = 1}^3 {{{\left| {f^i\left( {{\mathbf{w}_n}} \right) - q_n^i} \right|}^2}} \\
&\;\;\; \textup{s.t.} \;\; \textup{(\ref{omega})}, \textup{(\ref{schur})}.
\end{align}
\end{subequations}
It is observed that problem (\ref{omega_a}) is an SDP problem, which can be solved by convex optimization techniques \cite{grant2014cvx}.

Given ${\mathbf{\Omega}_n}$ and ${\mathbf{q}_n}$, the sub-problem of optimizing $\mathbf{w}_n$ is given by
\begin{subequations}  \label{w1}
\begin{align}
&\min \limits_{{\mathbf{w}_n}} \; -\sum\limits_{k = 0}^{K - 1} R_{u,k}^n + \frac{1}{{2\rho}} \sum\limits_{i = 1}^3 {{{\left| {f^i\left( {{\mathbf{w}_n}} \right) - q_n^i} \right|}^2}}  \label{obj3} \\
&\;\; \textup{s.t.} \;\, \textup{(\ref{pow})}.
\end{align}
\end{subequations}
Note that $ -\sum\limits_{k = 0}^{K - 1} R_{u,k}^n$ in (\ref{obj3}) is concave with respect to $\mathbf{w}_{u,k}^n$, while $\sum\limits_{i=1}^3 {{{\left| {f^i\left( {{\mathbf{w}_n}} \right) - q_n^i} \right|}^2}}$ is a non-convex quartic function with respect to $\mathbf{w}_n$.
Next, we transform both of them into convex forms, respectively.
Based on (\ref{rate}), by utilizing the SCA method \cite{MM7547360}, a convex surrogate function of ${\left| {{{( {\mathbf{h}_{u,k}^n} )}^H}\mathbf{w}_{u,k}^n} \right|^2}$ in $R_{u,k}^n$ at a given point ${\mathbf{w}_{u,k}^{n,(\gamma)}}$ is derived as
$2{\cal R}\{ {{( {\mathbf{w}_{u,k}^{n,(\gamma)}} )^H}\mathbf{H}_{u,k}^n \mathbf{w}_{u,k}^n} \} + {( {\mathbf{w}_{u,k}^{n,(\gamma)}} )^H} \mathbf{H}_{u,k}^n \mathbf{w}_{u,k}^{n,(\gamma)} - 2{\cal R}\{ {{( {\mathbf{w}_{u,k}^{n,(\gamma)}} )^H} \mathbf{H}_{u,k}^n \mathbf{w}_{u,k}^{n,(\gamma)}} \}$.
Hence, $-\sum\limits_{k = 0}^{K - 1} R_{u,k}^n$ is transformed into a convex function as
\begin{align}
&g_1 \left( {{\mathbf{w}_n}} \right) =  \nonumber  \\
& - \sum\limits_{k = 0}^{K - 1} {\textup{log}{_2}\left( {1 + 2{\cal R}\left\{ {{( {\mathbf{w}_{u,k}^{n,(\gamma)}} )^H}\mathbf{H}_{u,k}^n \mathbf{w}_{u,k}^n} \right\}/\sigma_{c}^2} \right)}.
\end{align}
Then, ${{{\left| {f^i\left( {{\mathbf{w}_n}} \right) - q_n^i} \right|}^2}}$ of the penalty term in (\ref{obj3}) is expanded as (\ref{penal_exp}) at the top of the next page, where $\tilde q_n^i = \upsilon_n^i - q_n^i$.
\begin{figure*}
\begin{minipage}{\textwidth}
\begin{align}  \label{penal_exp}
\vspace{0.01pt}
{\left| {f_n^i\left( {{\mathbf{w}_n}} \right) - q_n^i} \right|^2} = \sum\limits_{m = 1}^M \sum\limits_{k = 0}^{K-1} \sum\limits_{l = 0}^{L-1} \left\{ { \sigma_{{\theta _m}}^{- 4}{{\left| {{( {\mathbf{w}_{m,k}^n} )^H} \tilde{\cal A}_{m,k,l}^{n,i}\mathbf{w}_{m,k}^n} \right|}^2}} + 2{\sigma_{{\theta _m}}^{-2}{\cal R}\left\{ {{{\left( {\tilde q_n^i} \right)}^*}{( {\mathbf{w}_{m,k}^n} )^H}\tilde{\cal A}_{m,k,l}^{n,i} \mathbf{w}_{m,k}^n} \right\} } \right\}+ {\left| {\tilde q_n^i} \right|^2},
\vspace{0.01pt}
\end{align}
\end{minipage}
\end{figure*}
Note that the sub-terms of ${\left| {f^i\left( {{\mathbf{w}_n}} \right) - q_n^i} \right|^2}$ are non-convex.
Then, we leverage the SCA method to construct convex surrogate functions \cite{MM7547360}.
In specific, since $\mathrm{Tr} \left( \mathbf{ABCD} \right) = {\left( {\mathrm{vec}\left( {{\mathbf{D}^H}} \right)} \right)^H}\left( {{\mathbf{C}^T} \otimes \mathbf{A}} \right)\mathrm{vec}\left( \mathbf{B} \right)$, the sub-term ${\left| {{( {\mathbf{w}_{m,k}^n} )^H}\tilde{\cal A}_{m,k,l}^{n\,i}\mathbf{w}_{m,k}^n} \right|^2}$ in (\ref{penal_exp}) is reformulated as
${( {\mathbf{\bar {w}}_{m,k}^n} )^H}  \mathbf{\bar{\cal A}}_{m,k,l}^{n,i} \mathbf{\bar {w}}_{m,k}^n$,
where $\mathbf{\bar {w}}_{m,k}^n = \textup{vec}\left( {\mathbf{w}_{m,k}^n{( {\mathbf{w}_{m,k}^n} )^H}} \right)$
and $\mathbf{\bar{\cal A}}_{m,k,l}^{n,i} = {( {\tilde{\cal A}_{m,k,l}^{n,i}} )^*} \otimes \tilde{\cal A}_{m,k,l}^{n,i}$.
Then, the SCA method \cite{MM7547360} is adopted to construct the surrogate function of
${( {\mathbf{\bar {w}}_{m,k}^n} )^H} \mathbf{\bar{\cal A}}_{m,k,l}^{n,i} \mathbf{\bar {w}}_{m,k}^n$ as
$2{\cal R}\{ {\mathbf{a}_{m,k,l}^{n,i,(\gamma)}\mathbf{\bar {w}}_{m,k}^n} \} - 2{\cal R}\{ {\mathbf{a}_{m,k,l}^{n,i,(\gamma)}\mathbf{\bar {w}}_{m,k}^{n,(\gamma)}} \} + \mathbf{a}_{m,k,l}^{n,i,(\gamma)}\mathbf{\bar {w}}_{m,k}^{n,(\gamma)}$,
where $\mathbf{a}_{m,k,l}^{n,i,(\gamma)} = {( {\mathbf{\bar {w}}_{m,k}^{n,(\gamma)}} )^H} \mathbf{\bar{\cal A}}_{m,k,l}^{n,i}$.
By reshaping $\mathbf{a}_{m,k,l}^{n,i,(\gamma)}$ into a low-dimensional matrix, we obtain
$2{\cal R}\{ {\mathbf{a}_{m,k,l}^{n,i,(\gamma)}\mathbf{\bar {w}}_{m,k}^n} \}$ $ = 2{\cal R}\{ {{( {\mathbf{w}_{m,k}^n} )^H}\mathbf{\Xi}_{m,k,l}^{n,i,(\gamma)}\mathbf{w}_{m,k}^n} \}$, which is further equal to $2{\left| {{( {\mathbf{w}_{m,k}^{n,(\gamma)}} )^H}{{( {\tilde{\cal A}_{m,k,l}^{n,i}} )}^*}\mathbf{w}_{m,k}^n} \right|^2}$,
with ${\mathbf{\Xi}}_{m,k,l}^{n,i,(\gamma)} = {( {\tilde{\cal A}_{m,k,l}^{n,i}} )^T}\mathbf{w}_{m,k}^{n,(\gamma)}{( {\mathbf{w}_{m,k}^{n,(\gamma)}} )^H}{ ( {\tilde{\cal A}_{m,k,l}^{n,i}} )^*}$.
Then, the sub-item $2 \sigma_{{\theta_m}}^{-2} {{\cal R}\{ {{{\left( {\tilde q_n^i} \right)}^*}{( {\mathbf{w}_{m,k}^n} )^H}\tilde{\cal A}_{m,k,l}^{n,i}\mathbf{w}_{m,k}^n} \}}$ in (\ref{penal_exp}) is equivalently rewritten as ${\sigma_{{\theta_m}}^{-2}{( {\mathbf{w}_{m,k}^n} )^H}\mathbf{\Upsilon}_{m,k,l}^{n,i}\mathbf{w}_{m,k}^n}$,
where $\mathbf{\Upsilon}_{m,k,l}^{n,i} = {\left( {\tilde q_n^i} \right)^*}\tilde{\cal A}_{m,k,l}^{n,i} + \tilde q_n^i{( {\tilde{\cal A}_{m,k,l}^{n,i}} )^H}$.
Moreover, by employing the SCA method \cite{MM7547360}, the surrogate function of ${( {\mathbf{w}_{m,k}^n} )^H}\mathbf{\Upsilon}_{m,k,l}^{n,i}\mathbf{w}_{m,k}^n$  is constructed as
$2{\cal R}\{ {{( {\mathbf{w}_{m,k}^{n,(\gamma)}} )^H}\mathbf{\Upsilon}_{m,k,l}^{n,i}\mathbf{w}_{m,k}^n} \} + {( {\mathbf{w}_{m,k}^{n,(\gamma)}} )^H}\mathbf{\Upsilon}_{m,k,l}^{n,i}\mathbf{w}_{m,k}^{n,(\gamma)} -2{\cal R}\{ {{( {\mathbf{w}_{m,k}^{n,(\gamma)}} )^H}\mathbf{\Upsilon}_{m,k,l}^{n,i}\mathbf{w}_m^{n,(\gamma)}} \}$.
Therefore, ${{{\left| {f^i\left( {{\mathbf{w}_n}} \right) - q_n^i} \right|}^2}}$ is converted into the convex function in (\ref{f2}) at the top of the next page.
\begin{figure*}
\begin{minipage}{\textwidth}
\begin{align}  \label{f2}
\vspace{0.01pt}
g_2^i \left( \mathbf{w}_n,\mathbf{q}_n \right)= 2 \sum\limits_{m = 1}^M \sum\limits_{k = 0}^{K-1} \sum\limits_{l = 0}^{L-1} { \left\{ {\sigma _{{\theta_m}}^{-4}{{\left| {{( {\mathbf{w}_{m,k}^{n,(\gamma)}} )^H}{{( {\tilde{\cal A}_{m,k,l}^{n,i}} )}^*}\mathbf{w}_{m,k}^n} \right|}^2} + \sigma _{{\theta_m}}^{-2}{\cal R}\left\{ {{( {\mathbf{w}_{m,k}^{n,(\gamma)}} )^H}\mathbf{\Upsilon}_{m,k,l}^{n,i}\mathbf{w}_{m,k}^n} \right\}} \right\}} + {\left| {\tilde q_n^i} \right|^2},
\vspace{0.01pt}
\end{align}
\end{minipage}
\end{figure*}
Then, problem (\ref{w1}) is reformulated as
\begin{subequations}  \label{problem_wn}
\begin{align}
&\min \limits_{{\mathbf{w}_n}} \; {g_1}\left( {{\mathbf{w}_n}} \right) + \frac{1}{{2\rho}} \sum\limits_{i = 1}^3 {g_2^i}\left( {{\mathbf{w}_n},{\mathbf{q}_n}} \right)\\
&\;\, \textup{s.t.} \;\; \textup{(\ref{pow})},
\end{align}
\end{subequations}
which is convex and thus can be tackled by convex optimization techniques such as CVX via the interior-point method \cite{grant2014cvx}.

\subsubsection{Outer Layer Update}
Recall that the penalty factor $\rho$ is initialized by a large value and gradually decreased in the outer layer.
Hence, $\rho$ in the $\lambda$-th iteration is updated as $\rho^{\left( \lambda \right)}= \xi  \rho^{\left( \lambda-1 \right)}$, where $0<\xi <1$ is the updated step size. From experiments, $\xi$ can be set between 0.7 and 0.9 to achieve an effective trade-off between performance and complexity.

\subsubsection{Overall Algorithm}
To verify whether the solution of problem (\ref{add_penalty}) violates the equality constraints in (\ref{penalty}) or not, the termination indicator is defined as
\begin{align}
\kappa = \max \left\{ {{{\left| {f^i\left( {{\mathbf{w}_n}} \right) - q_n^i} \right|}^2},\forall i \in {\cal I}} \right\}.
\end{align}
When $\kappa$ is smaller than the predefined value, the equality constraints in (\ref{penalty}) are satisfied and the algorithm is terminated.
The details of the proposed penalty-based algorithm are summarized in Algorithm 1.
\begin{algorithm}[t]
\label{algorithm1}
\caption{Penalty-based Algorithm for Solving (\ref{add_penalty}).}
\begin{enumerate}
\item \textbf{Input}: $\mathbf{w}_n^{\left( 0 \right)}$, $\rho^{\left( \lambda \right)}$, iteration index $\lambda=1$, step size $\xi$.
\item \textbf{Output}: $\mathbf{w}_n$.
\item \textbf{while} no convergence \textbf{do}
\begin{enumerate}
\item \textbf{while} no convergence \textbf{do}
\begin{enumerate}
\item Update $ \mathbf{\Omega}_n $ and $ {{\mathbf{q}_n}} $ by solving problem (\ref{omega_a}).
\item Update $\mathbf{w}_n$ by solving problem (\ref{problem_wn}).
\end{enumerate}
\item \textbf{end while}
\item Update $\lambda = \lambda+1$.
\item Update ${{\rho }^{\left( \lambda \right)}}= \xi {{\rho }^{\left( \lambda - 1 \right)}}$.
\end{enumerate}
\item \textbf{end while}
\end{enumerate}
\end{algorithm}

As the penalty coefficient $\rho$ decreases, the equality constraints in (\ref{penalty}) are ultimately guaranteed.
In each inner iteration, the sub-problem of $ \mathbf{\Omega}_n$ and ${{\mathbf{a}_n}}$, and the sub-problem of $\mathbf{w}_n$ are solved alternately with the other fixed.
The former achieves the optimal solution, while the latter obtains a non-increasing objective value of problem (\ref{problem_wn}) through the SCA method.
Thereby, the objective function of problem (\ref{add_penalty}) is non-decreasing over the inner iterations.
In addition, due to the PC-CRLB requirement and the transmit power budget of the BSs, the objective function is upper-bounded by a finite value.
Hence, the solution of Algorithm 1 is guaranteed to converge.

\subsection{Complexity Analysis}
Based on the interior-point method, the computational complexity for addressing the SDP problem (SDR1) is $\mathcal{O}\left( \log \frac{1}{\zeta }\left( KM+2 \right)^6 N_t ^6 \right)$ \cite{hanif2014computationally}, where $\zeta$ is the convergence accuracy. Hence, the computational complexity of the SDR-based algorithm is $\mathcal{O}\left( \log \frac{1}{\zeta }\left( KM+2 \right)^6 N_t^6 \right)$.
It is observed that the computational complexity of the penalty-based algorithm is determined by solving problem (\ref{problem_wn}) for updating $\mathbf{w}_n$.
This results in a complexity of ${\cal O}\left( I_1{\sqrt {N_t + M} {N_t^3}} \right)$ using the interior-point method, where $I_1$ is the number of iterations for the SCA method.
Thus, the overall computational complexity of the penalty-based algorithm is ${\cal O}\left\{ {{I_2}{I_3}{{I_1}\sqrt {N_t + M} {N_t^3}}} \right\}$, where $I_2$ and $I_3$ are the number of iterations for the inner layer and outer layer to achieve convergence, respectively.
Notably, compared to the SDR-based algorithm, the penalty-based algorithm achieves lower computational complexity\footnote{The interior-point method is used to validate the performance of the proposed algorithms, providing exact and stable solutions. By leveraging this method, the penalty-based algorithm achieves a significantly lower computational complexity than the SDR-based algorithm. For scenarios with limited computational resources, methods such as the gradient projection descent method \cite{liu2009efficient} can be further applied to inexactly solve problem (\ref{problem_wn}), bypassing the interior-point method and reducing complexity. Further exploration of complexity reduction techniques is left for future work.}.

\section{Simulation Results}
In this section, we present simulation results to evaluate the performance of the proposed schemes in the MIMO-OFDM networked ISAC system.
In specific, we consider $M=3$ BSs located at ${\mathbf{b}_1} = {\left[ {-50 \; \textup{m},0 } \right]^T}$ for BS 1, ${\mathbf{b}_2} = {\left[ {50 \; \textup{m},0 } \right]^T}$ for BS 2, and ${\mathbf{b}_3} = {\left[ {0 ,-50 \; \textup{m}} \right]^T}$ for BS 3, with carrier frequencies of $28$ $\textup{GHz}$, $29$ $\textup{GHz}$, and $30$ $\textup{GHz}$, respectively.
Without loss of generality, each BS is equipped with an equal number of transmit and receive antennas, i.e., $N_t=N_r=N=4$.
Unless specified otherwise, the transmit power budget of each BS is ${P_m}=40$ $\textup{dBm}$, $\forall m \in {\cal M}$ and the PC-CRLB threshold is $\eta=0.01$.
The initial state of the target is ${\mathbf{x}_0} = {\left[ {40 \; \textup{m},0,20 \; \textup{m/s},0} \right]^T}$ and state evolution noises are ${\sigma_{{p_x}}} = {\sigma _{{p_y}}} = 0.02$ m, ${\sigma _{{v_x}}} = {\sigma _{{v_y}}} = 0.01$ $\textup{m/s}$.
The TTS duration is $\Delta T=0.02$ $\textup{s}$, and the maximum time duration of interest is $T=4$ $\textup{s}$.
It is assumed that the RCS of the target remains constant within the short duration, i.e., $\beta^n=\beta^{n-1}=0.2+0.2j$.
Besides, the employed distance-dependent path-loss model is $PL( \tilde d ) = {C_0}{( {\tilde d / {\tilde d_0}} )^{ - \alpha }}$, where $C_0$ is the path loss at the reference distance $\tilde d_0$, $\alpha$ is the path-loss exponent, and $\tilde d$ is the link distance.
We set the path-loss exponents of all links to $\alpha=2$.
The remaining system parameters are set as $\Delta f = 120$ kHz, $K=4096$, $B = 491.52$ MHz, $T_e = 8.33$ $\mu s$, $T_{c} = 0.59$ $\mu s$, $L=256$, $c=3 \times 10^8$ $\textup{m/s}$, $\xi=0.8$, $C_0=30$ $\textup{dBm}$, and $\tilde d_0=1$ $\textup{m}$.
Additionally, the power spectrum density of AWGN at each BS/device is $-167$ dBm/Hz, with the bandwidth $B = 491.52$ MHz, thus yielding $\sigma_c^2=\sigma_m^2=-80$ $\textup{dBm}$.

\subsection{Tracking Performance of the Proposed Schemes}
\begin{figure}[t]
\noindent \centering{}\includegraphics[scale=0.46]{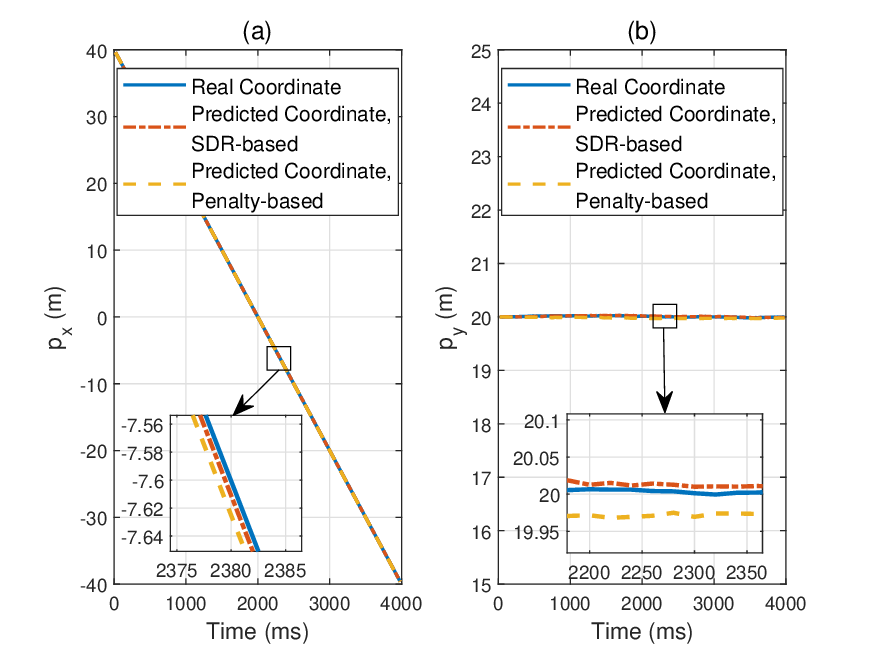}\caption{Performance of target position tracking on the (a) x-axis; (b) y-axis.}\label{x-y_epoch}
\end{figure}

\begin{figure}[t]
\noindent \centering{}\includegraphics[scale=0.46]{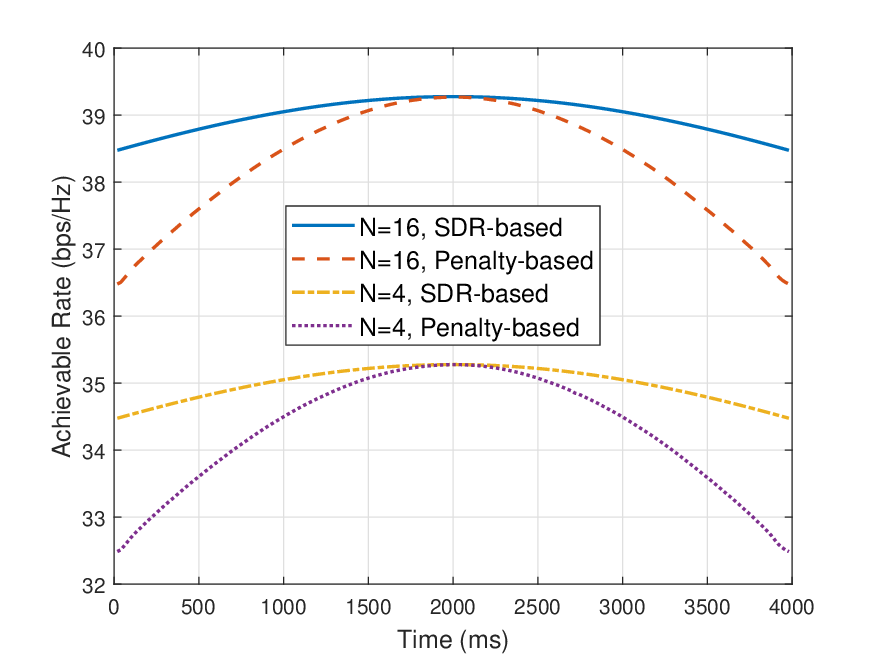}\caption{Achievable rate versus the time instant within the maximum time duration of interest $T$.}\label{rate_epoch_antenna}
\end{figure}

\begin{figure}[t]
\noindent \centering{}\includegraphics[scale=0.46]{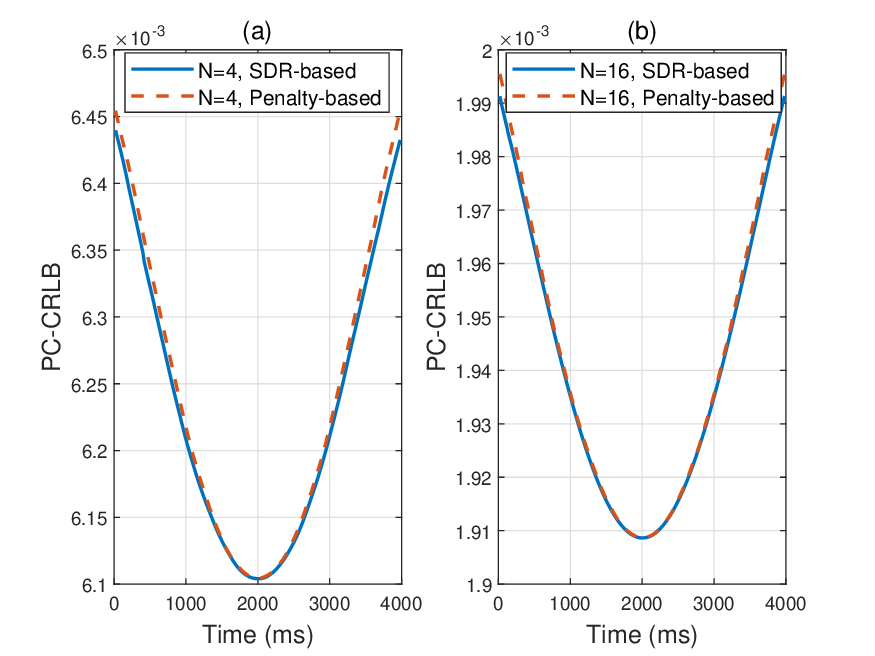}\caption{Achieved PC-CRLB versus the time instant within the maximum time duration of interest $T$ under (a) $N=4$; (b) $N=16$. }\label{CRLB_epoch}
\end{figure}

\begin{figure}[t]
\noindent \centering{}\includegraphics[scale=0.46]{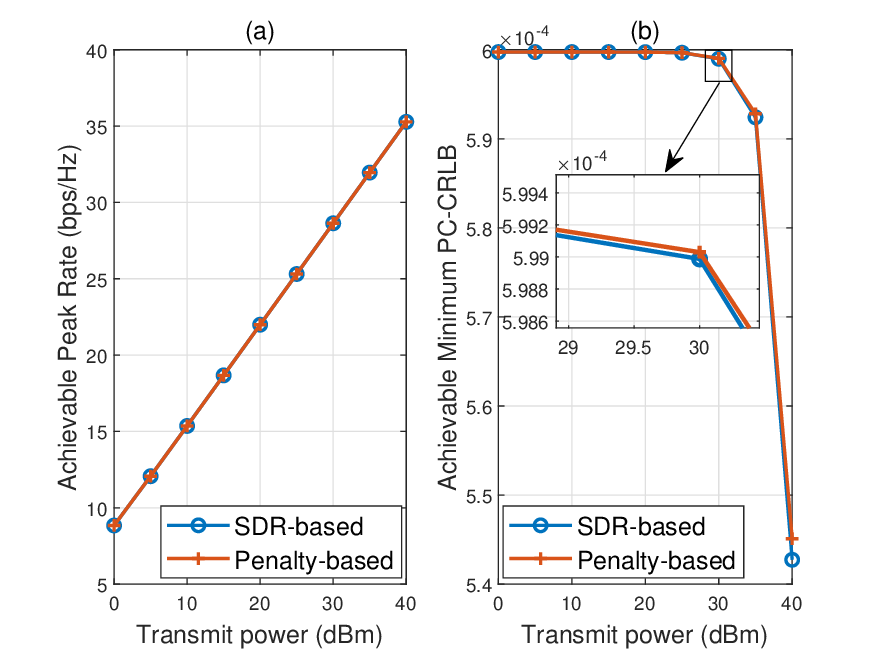}\caption{(a) Achievable peak rate versus the transmit power; (b) Achieved minimum PC-CRLB versus the transmit power. }\label{rate-CRLB_power}
\end{figure}

\begin{figure}[t]
\noindent \centering{}\includegraphics[scale=0.46]{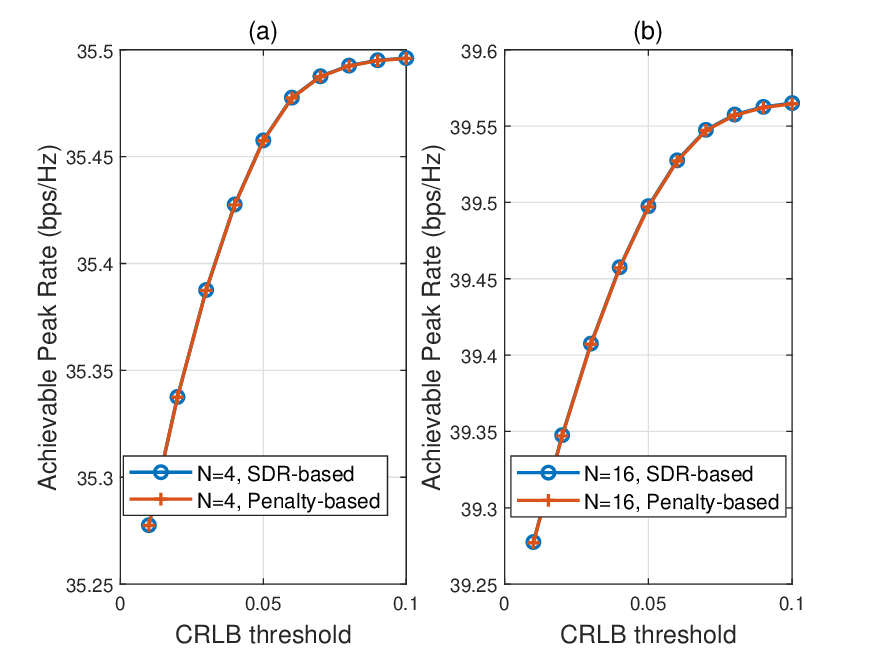}\caption{Communication-sensing tradeoff under (a) $N=4$; (b) $N=16$. }\label{tradeoff}
\end{figure}

\begin{figure}[t]
\noindent \centering{}\includegraphics[scale=0.46]{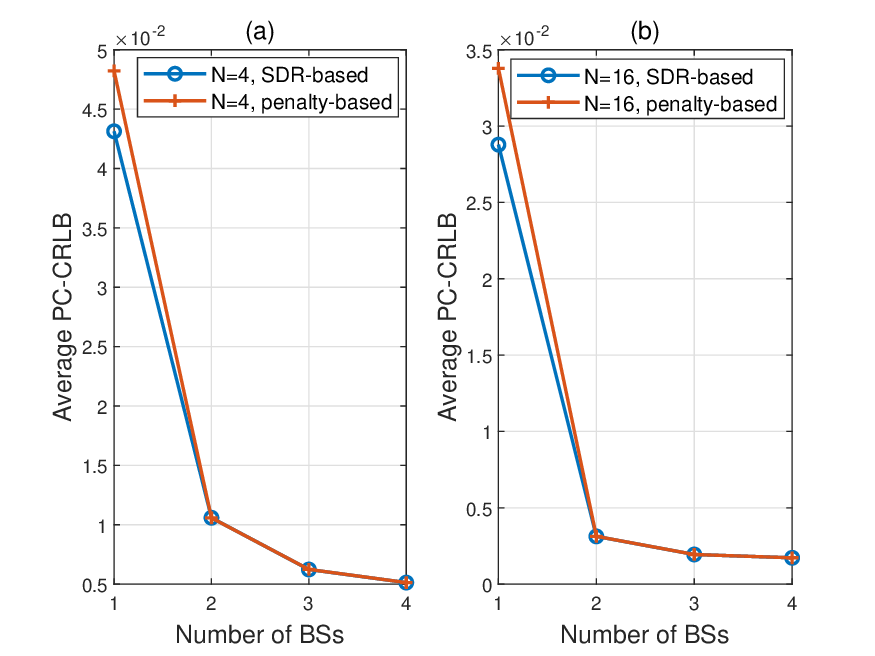}\caption{Achieved average PC-CRLB versus the number of BSs under (a) $N=4$; (b) $N=16$. }\label{numBS}
\end{figure}

\begin{figure}[t]
\noindent \centering{}\includegraphics[scale=0.46]{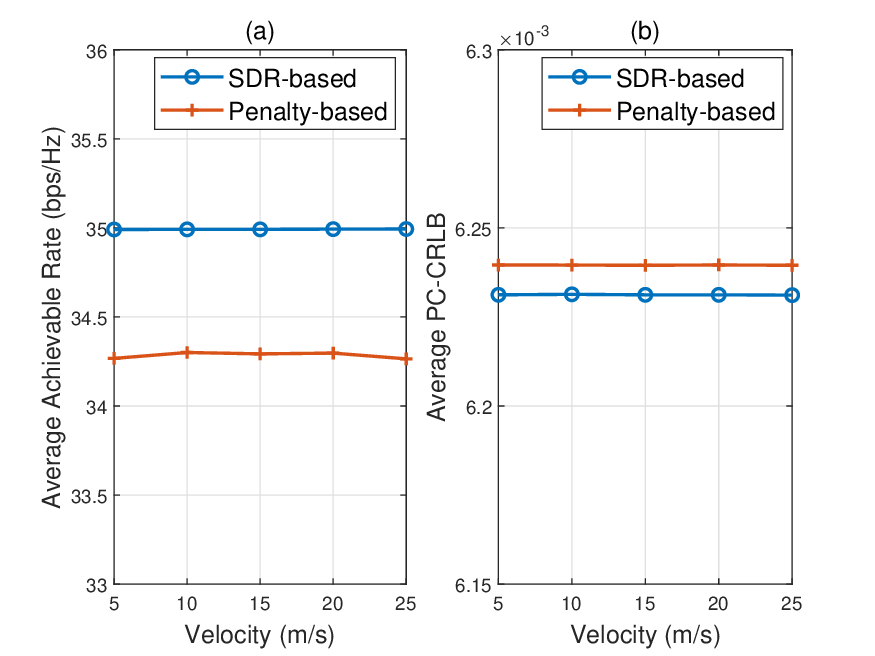}\caption{(a) Average achievable rate versus the velocity; (b) Achieved average PC-CRLB versus the velocity.}\label{rate-CRLB_vx}
\end{figure}

In this subsection, based on the proposed SDR-based (``SDR-based") and penalty-based (``penalty-based") predictive beamforming design algorithms, respectively, we evaluate the tracking performance of the developed cooperative sensing-assisted predictive beam tracking schemes.

Fig. \ref{x-y_epoch} shows the tracking results presented on the x-axis and y-axis, including the true coordinates of the target, as well as the predicted coordinates obtained by the cooperative sensing-assisted beam tracking schemes enabled by the SDR-based and the penalty-based predictive beamforming strategies, respectively.
It is observed that the predicted coordinates of the proposed schemes closely match the true target coordinates, with the SDR-based scheme yielding more accurate predictions.
This is because compared to the suboptimal solution obtained by the penalty-based scheme with lower complexity, the SDR-based scheme achieves the optimum for predictive beamforming with higher accuracy.
Overall, both schemes maintain high tracking accuracy throughout the process.

To validate the impact of target motion on communication performance, Fig. \ref{rate_epoch_antenna} illustrates the achievable rates of the device for the proposed schemes versus the time instant within the maximum time duration with different antenna settings, i.e., $N=4$ and $N=16$, respectively.
We observe that the achievable rates achieved by all schemes increase first and then decrease as the target moves.
This is because as the target gradually moves closer to BS 1 and farther away from BS 2, the beamforming gain of BS 1 increases while that of BS 2 decreases, leading to mutual cancellation. Meanwhile, the gain attained by BS 3 providing communication services increases initially and then decreases as the target moves, causing a similar trend for achievable rates.
Notably, at the peak array gain provided by BS 3, the penalty-based scheme achieves the same performance as the SDR-based scheme. This phenomenon indicates that the array gain compensates for the accuracy loss in low-complexity penalty-based solutions.
Additionally, under different antenna settings, since the SDR-based algorithm provides the optimal predicting beamforming, the SDR-based scheme achieves superior communication performance.
Besides, BSs with more antennas achieve higher achievable rates due to their ability to provide greater array gain.

To evaluate the impact of target motion on sensing performance, we illustrate the achieved PC-CRLB of the proposed schemes versus the time instant in Fig. \ref{CRLB_epoch}, with $N=4$ and $N=16$, respectively. Note that the achieved PC-CRLB under different settings decreases first and then increases as the target moves, following the same reasoning as Fig. \ref{rate_epoch_antenna}.
Moreover, the SDR-based scheme and the penalty-based scheme achieve the same PC-CRLB at the maximum array gain, which is consistent with their identical peak rates in Fig. \ref{rate_epoch_antenna}.
Besides, as seen in Fig. \ref{CRLB_epoch} (a) and Fig. \ref{CRLB_epoch} (b), BSs equipped with more antennas achieve lower PC-CRLB values.
Furthermore, it is clear that during the target tracking process, the achieved PC-CRLB satisfies the pre-defined threshold, i,e., $\eta=0.01$, effectively demonstrating the validity of the proposed schemes.

Fig. \ref{rate-CRLB_power} shows the communication performance measured by the achievable peak rate and sensing performance measured by the achievable minimum PC-CRLB during the tracking process versus the transmit power budget $P_m,\forall m \in {\cal M}$, respectively.
From Fig. \ref{rate-CRLB_power} (a), note that an increased transmit power of the BSs results in higher achievable rates of the device.
In addition, it is essential to point out that the SDR-based design achieves nearly the same achievable peak rate as the penalty-based design, consistent with the observations in Fig. \ref{rate_epoch_antenna}. In Fig. \ref{rate-CRLB_power} (b), similar trends to Fig. \ref{rate-CRLB_power} (a) are observed at lower transmit power, while as the transmit power increases, the SDR-based scheme achieves slightly better performance than the penalty-based scheme.

In Fig. \ref{tradeoff}, we study the performance tradeoff between communication and sensing, under $N=4$ and $N=16$, respectively.
It is observed that the achievable peak rates for communication during the tracking process gradually increase as the sensing requirements decrease. This is due to the fact that the objectives of performing sensing and communication tasks are not entirely consistent, leading to differences in the design of predictive beamformers to satisfy the PC-CRLB constraint for sensing and maximize the achievable rate for communication.
Additionally, from Fig. \ref{tradeoff} (a) and Fig. \ref{tradeoff} (b), one observes that setups with more antennas achieve higher communication
rates, which is consistent with the observations in Fig. \ref{rate_epoch_antenna}.

To demonstrate the effectiveness of multi-BS cooperative sensing, Fig. \ref{numBS} illustrates the impact of the number of BSs on sensing performance, measured by the average PC-CRLB during tracking, with $N=4$ and $N=16$, respectively.
In the single-BS setup, only BS 3 at ${\mathbf{b}_3} = {\left[ {0, -50 \; \textup{m}} \right]^T}$ is considered; in the two-BS setup, BS 3 and BS 4 at ${\mathbf{b}_4} = {\left[ {0, 50 \; \textup{m}} \right]^T}$ are included; in the three-BS setup, BS 1, 2, and 3 are used; and in the four-BS setup, BS 4 is added to the three-BS configuration.
As the number of BSs increases, the average PC-CRLB gradually decreases, owing to the richer observations provided by the BSs, enhancing target estimation accuracy. This highlights the importance of multi-BS cooperative sensing.
Moreover, the SDR-based scheme achieves a lower average PC-CRLB than the penalty-based scheme. Notably, the performance gap between the two schemes becomes smaller as the number of BSs increases, as more BSs provide richer target information and greater array gains, compensating for the accuracy loss in penalty-based solutions. Besides, as seen in Fig. \ref{numBS} (a) and Fig. \ref{numBS} (b), BSs with more antennas achieve a lower average PC-CRLB, which is consistent with the results in Fig. \ref{CRLB_epoch}.

To verify the performance of the proposed schemes under high mobility, Fig. \ref{rate-CRLB_vx} (a) and Fig. \ref{rate-CRLB_vx} (b) show the average achievable rate and the achieved average PC-CRLB during the tracking process versus the target velocity, respectively.
It should be noted that the target follows the same trajectory at different velocities.
One can observe that both the SDR-based scheme and the penalty-based scheme exhibit strong robustness to velocity in terms of achievable rate and achieved PC-CRLB, indicating that our proposed schemes can accurately track the target across different velocities.
Additionally, from Fig. \ref{rate-CRLB_vx} (a), it is observed that the SDR-based design achieves higher achievable rates than the penalty-based design, as it provides a global optimum for predictive beamforming.
Moreover, as shown in Fig. \ref{rate-CRLB_vx} (b), the SDR-based design achieves a lower PC-CRLB, which is consistent with the observations in Fig. \ref{CRLB_epoch}.

\subsection{Performance Comparison with the Benchmark Schemes}
\begin{figure}[t]
\noindent \centering{}\includegraphics[scale=0.46]{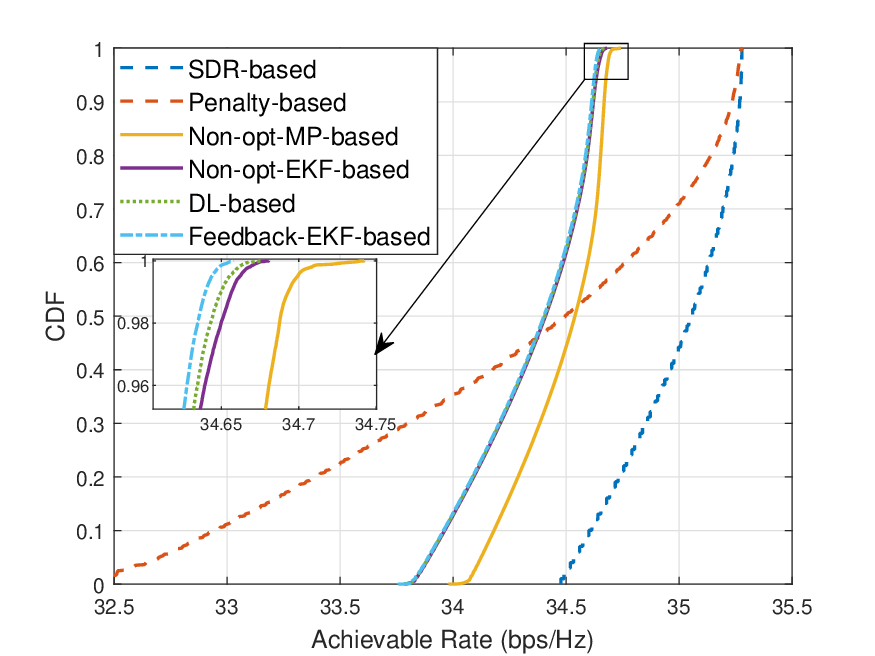}\caption{CDF of the achievable rate.}\label{CDF_rate}
\end{figure}

\begin{figure}[t]
\noindent \centering{}\includegraphics[scale=0.46]{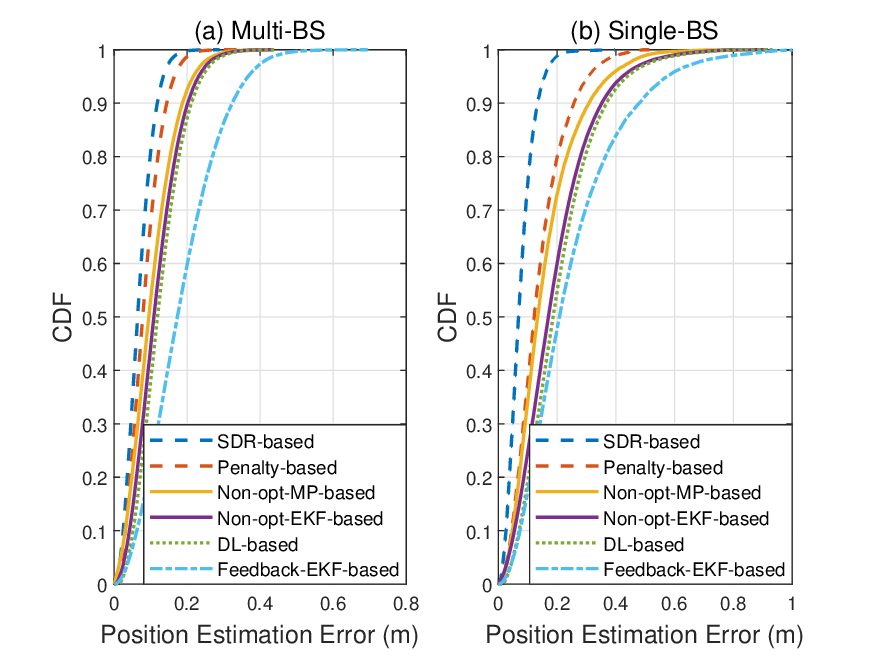}\caption{CDF of the position estimation error under (a) multi-BS scenario; (b) single-BS scenario.}\label{CDF_error}
\end{figure}

In this subsection, we compare the proposed schemes with the following benchmark schemes to verify the effectiveness:
\begin{itemize}
\item
Feedback-enabled EKF-based scheme such as \cite{liu2019ekf} (``Feedback-EKF-based"):
the BS sends a pilot to the target within each TTS, then the target estimates the angle and feeds it back for setting the beam of the next TTS by utilizing the EKF method.
\item
ISAC-enabled EKF-based predictive beam tracking scheme without beamforming optimization \cite{liu2020radar} (``Non-opt-EKF-based"):
the BS processes the echo signal to obtain the target parameters and then aligns the beam towards the predicted angle obtained from the EKF method, without optimizing the predictive beamforming.
\item
ISAC-enabled MP-based predictive beam tracking scheme without beamforming optimization \cite{yuan2020bayesian} (``Non-opt-MP-based"):
the BS processes the echo signal to obtain the target parameters and then aligns the beam towards the predicted angle obtained from the MP method, without optimizing the predictive beamforming.
\item
ISAC-enabled DL-based predictive beam tracking scheme \cite{zhang2024predictive} (``DL-based"):
the BS processes the echo signal to obtain the target parameters and then employs the DL-based method to map the historical channel to the predictive beamforming.
\end{itemize}
For a fair comparison, unless otherwise specified, the same state evolution model, measurement model, and system parameter settings are used for all schemes.

Fig. \ref{CDF_rate} shows the cumulative distribution functions (CDF) of the achievable rates for all schemes.
It is observed that the SDR-based scheme achieves the best communication performance, due to its ability to provide the optimal solution with high accuracy.
Moreover, the proposed SDR-based and penalty-based schemes outperform the benchmark schemes in terms of achievable peak rates, demonstrating the effectiveness of predictive beamforming optimization.
Besides, we observe that the penalty-based scheme yields relatively lower rates, especially when the target is far from BS 3 providing services, as indicated in Fig. \ref{rate_epoch_antenna}.
This is due to its suboptimal solution for predictive beamforming, which results in accuracy loss, particularly in regions with limited array gain.
Conversely, when the target moves closer to BS 3 and benefits from higher array gain, note that the penalty-based scheme performs similarly to the SDR-based scheme.
These results indicate that the penalty-based scheme can be employed when the target is near the serving BS, as it achieves superior performance with lower computational complexity.
In addition, it is worth pointing out that compared to the proposed schemes, the DL-based
scheme is expected to offer a significant advantage in computational complexity when the BSs are equipped with extremely large-scale antenna arrays \cite{ma2022deep}.
Furthermore, one observes that the feedback-EKF-based scheme achieves slightly lower rates compared to the non-optimization schemes and the DL-based scheme, without extremely low rates. This differs from the observations in \cite{liu2020radar}, which focuses on a single-BS scenario where the feedback-EKF-based scheme results in some extremely low rates due to abrupt changes in the tracking angle as the target approaches the BS.
The difference arises because multi-BS cooperative sensing provides multiple observations, ensuring that the angles of the target relative to the BSs do not change suddenly and simultaneously, thus significantly reducing the misalignment probability.
Additionally, \cite{liu2020radar} shows that under slow angle variations, the feedback-EKF-based scheme achieves similar rate performance to the non-opt-EKF-based scheme, consistent with the findings in the multi-BS cooperative sensing scenario, where sudden angle variations are effectively avoided. These phenomena demonstrate the stability of multi-BS cooperative sensing.

In Fig. \ref{CDF_error}, we present the CDF of position estimation errors during tracking for all schemes under both multi-BS cooperative sensing and single-BS sensing, respectively. In the single-BS setup, only BS 3 at ${\mathbf{b}_3} = {\left[ {0 ,-50 \; \textup{m}} \right]^T}$ is considered.
It is observed that compared to single-BS sensing, multi-BS cooperative sensing significantly enhances target estimation accuracy by providing several observations of the target, resulting in a notably lower maximum estimation error.
This indicates that multi-BS cooperation mitigates the impact of signal attenuation and blind zones at a single BS, thereby enhancing system robustness.
As shown in Fig. \ref{CDF_error} (a) and Fig. \ref{CDF_error} (b), we observe that the proposed schemes achieve superior estimation performance compared to the non-optimization schemes and the DL-based scheme. This demonstrates the effectiveness of using the PC-CRLB as the sensing performance metric for optimizing the predictive beamforming and the corresponding proposed algorithms.
In addition, one can observe that the feedback-EKF-based scheme exhibits larger estimation errors.
This is because, unlike the ISAC-enabled schemes, the feedback-EKF-based scheme only uses a small number of pilot symbols for target estimation, which may lead to partial loss of target information.
Overall, the proposed schemes achieve better target tracking performance compared to the benchmark schemes.

\section{Conclusion}
In this paper, we investigated a MIMO-OFDM networked ISAC system and proposed a cooperative sensing-assisted predictive beam tracking scheme to track the mobile device as a moving target.
Specifically, a cooperative sensing-assisted target tracking approach by utilizing the EKF technique was developed to estimate and predict the target parameters.
Then, based on the target prediction, we proposed a predictive beamforming design scheme within individual TTS to facilitate achieving predictive beam tracking for continuous TTSs.
In particular, we derived the achievable rate for communication and the PC-CRLB for sensing, and then formulated the rate maximization problem by designing the predictive beamforming of the BSs, while satisfying the PC-CRLB constraint.
To tackle the resulting non-convex problem, an SDR-based algorithm and a penalty-based algorithm were proposed to obtain the optimal solution and low-complexity solution, respectively.
Finally, simulation results showed that the proposed schemes outperformed the benchmark schemes, and the SDR-based design achieved excellent communication performance.
The results also showed that compared to single-BS sensing, multi-BS cooperative sensing achieved better estimation performance.

\section*{Appendix}
\section*{Derivation of the Predicted Conditional Fisher Information Matrices}

${\mathbf{J}_{{\mathbf{p}_n}{\mathbf{p}_n}}}$ can be expanded as
${\mathbf{J}_{{\mathbf{p}_n}{\mathbf{p}_n}}} = \left[ {\begin{array}{*{20}{c}}
{{J_{{p_x}{p_x}}^n}}&{{J_{{p_x}{p_y}}^n}}\\
{{J_{{p_x}{p_y}}^n}}&{{J_{{p_y}{p_y}}^n}}
\end{array}} \right]$.
Based on the derivations in (\ref{G_m}), (\ref{OP}), (\ref{JP}), and (\ref{J_O}), the elements of ${\mathbf{J}_{{\mathbf{p}_n}{\mathbf{p}_n}}}$ are derived as
\begin{align}
& {J_{{p_x}{p_x}}^n}={\sum\limits_{m=1}^M \sum\limits_{k=0}^{K-1} \sum\limits_{l=0}^{L-1} {\sigma_{{\theta_m}}^{-2}\mathrm{Tr}\left( { \tilde{\cal A}_{m,k,l}^{n,1} \mathbf{w}_{m,k}^n{{( {\mathbf{w}_{m,k}^n} )}^H}} \right)} + \upsilon_n^1 },  \label{a1} \\
& {J_{{p_x}{p_y}}^n}={\sum\limits_{m=1}^M \sum\limits_{k=0}^{K-1} \sum\limits_{l=0}^{L-1} {\sigma_{{\theta_m}}^{-2}\mathrm{Tr}\left( { \tilde{\cal A}_{m,k,l}^{n,2} \mathbf{w}_{m,k}^n{{( {\mathbf{w}_{m,k}^n} )}^H}} \right)} + \upsilon_n^2 }, \\
& {J_{{p_y}{p_y}}^n}={\sum\limits_{m=1}^M \sum\limits_{k=0}^{K-1} \sum\limits_{l=0}^{L-1} {\sigma_{{\theta_m}}^{-2}\mathrm{Tr}\left( { \tilde{\cal A}_{m,k,l}^{n,3} \mathbf{w}_{m,k}^n{{( {\mathbf{w}_{m,k}^n} )}^H}} \right)} + \upsilon_n^3 },
\end{align}
where
$\tilde{\cal A}_{m,k,l}^{n,1}={{( { {\cal A}_{m,k,l}^{y,n}} )}^H} {\cal A}_{m,k,l}^{y,n}$,
$\tilde{\cal A}_{m,k,l}^{n,2}=$ ${{( { {\cal A}_{m,k,l}^{y,n}} )}^H} {\cal A}_{m,k,l}^{x,n}$,
and $\tilde{\cal A}_{m,k,l}^{n,3}={{( { {\cal A}_{m,k,l}^{x,n}} )}^H} {\cal A}_{m,k,l}^{x,n}$,
with ${\cal A}_{m,k,l}^{x,n}$ and ${\cal A}_{m,k,l}^{y,n}$ calculated based on the state prediction $\mathbf{\hat d}_{n\left| {n - 1} \right.}$.
Furthermore, $\upsilon_n^1=\sum\limits_{m = 1}^M \bar a_m^{n,1} + {\varsigma_{11}^n}$,
$\upsilon_n^2=\sum\limits_{m = 1}^M \bar a_m^{n,2} + {\varsigma_{12}^n}$, and $\upsilon_n^3=\sum\limits_{m = 1}^M \bar a_m^{n,3} + {\varsigma_{22}^n}$,
where $\bar a_m^{n,1} = {\frac{{4\sigma_{{\tau_m}}^{-2}{{\left( {\ell_{x,m}^{n}} \right)}^2}}}{{{c^2}{{\left\| {{\mathbf{p}_{n\left| {n-1} \right.}} - {\mathbf{b}_m}} \right\|}^2}}}} + {\frac{{4{{ {f_m^2} }}\sigma_{{\mu_m}}^{-2}{{\left( {\ell_{y,m}^{n}} \right)}^2}{{\left( {\psi_{x,y,m}^{n}} \right)}^2}}}{{{c^2}{{\left\| {{\mathbf{p}_{n\left| {n-1} \right.}} - {\mathbf{b}_m}} \right\|}^6}}}}$,
$\bar a_m^{n,2} ={\frac{{4\sigma_{{\tau_m}}^{-2}\ell_{x,m}^{n}\ell_{y,m}^{n}}}{{{c^2}{{\left\| {{\mathbf{p}_{n\left| {n - 1} \right.}} - {\mathbf{b}_m}} \right\|}^2}}}} + {\frac{{ -4{{ {f_m^2} }}\sigma_{{\mu_m}}^{-2}\ell_{x,m}^{n}\ell_{y,m}^{n}{{\left( {\psi_{x,y,m}^{n}} \right)}^2}}}{{{c^2}{{\left\| {{\mathbf{p}_{n\left| {n-1} \right.}} - {\mathbf{b}_m}} \right\|}^6}}}}$,
and $\bar a_m^{n,3} = {\frac{{4\sigma_{{\tau_m}}^{-2}{{\left( {\ell_{y,m}^{n}} \right)}^2}}}{{{c^2}{{\left\| {{\mathbf{p}_{n\left| {n-1} \right.}} - {\mathbf{b}_m}} \right\|}^2}}}} + {\frac{{4{{{f_m^2}}}\sigma_{{\mu_m}}^{-2}{{\left( {\ell_{x,m}^{n}} \right)}^2}{{\left( {\psi_{x,y,m}^{n}} \right)}^2}}}{{{c^2}{{\left\| {{\mathbf{p}_{n\left| {n-1} \right.}} - {\mathbf{b}_m}} \right\|}^6}}}} $,
with $\ell_{x,m}^{n}$, $\ell_{y,m}^{n}$, and $\psi_{x,y,m}^{n}$ calculated based on the state prediction $\mathbf{\hat d}_{n\left| {n - 1} \right.}$.
Moreover, $\varsigma_{ij}^n$ denotes the $(i,j)$-th element of $\mathbf{M}_{n\left| {n - 1} \right.}^{-1}$.

Similarly, ${\mathbf{J}_{{\mathbf{p}_n}{\mathbf{v}_n}}}$ can be expanded as ${\mathbf{J}_{{\mathbf{p}_n}{\mathbf{v}_n}}} = \left[ {\begin{array}{*{20}{c}}
{{J_{{p_x}{v_x}}^n}}&{{J_{{p_x}{v_y}}^n}}\\
{{J_{{p_y}{v_x}}^n}}&{{J_{{p_y}{v_y}}^n}}
\end{array}} \right]$.
According to the derivations in (\ref{G_m}), (\ref{OP}), (\ref{JP}), and (\ref{J_O}), the elements of ${\mathbf{J}_{{\mathbf{p}_n}{\mathbf{v}_n}}}$ are calculated as
\begin{align}
& {J_{{p_x}{v_x}}^n} = {\sum\limits_{m = 1}^M {\frac{{4f_m^2\sigma_{{\mu_m}}^{- 2}\ell_{x,m}^{n}\ell_{y,m}^{n}\psi_{x,y,m}^{n}}}{{{c^2}{{\left\| {{\mathbf{p}_{n\left| {n - 1} \right.}} - {\mathbf{b}_m}} \right\|}^4}}}} + {\varsigma_{13}^n}}, \\
& {J_{{p_x}{v_y}}^n} = {\sum\limits_{m = 1}^M {\frac{{4f_m^2\sigma_{{\mu_m}}^{- 2}{{\left( {\ell_{y,m}^{n}} \right)}^2}\psi_{x,y,m}^{n}}}{{{c^2}{{\left\| {{\mathbf{p}_{n\left| {n - 1} \right.}} - {\mathbf{b}_m}} \right\|}^4}}}} + {\varsigma_{14}^n}}, \\ & {J_{{p_y}{v_x}}^n} = {\sum\limits_{m = 1}^M {\frac{{-4f_m^2\sigma_{{\mu _m}}^{- 2}{{\left( {\ell_{x,m}^{n}} \right)}^2}\psi_{x,y,m}^{n}}}{{{c^2}{{\left\| {{\mathbf{p}_{n\left| {n - 1} \right.}} - {\mathbf{b}_m}} \right\|}^4}}}} + {\varsigma_{23}^n}}, \\
& {J_{{p_y}{v_y}}^n} = {\sum\limits_{m = 1}^M {\frac{{-4f_m^2\sigma_{{\mu _m}}^{- 2}\ell_{x,m}^{n}\ell_{y,m}^{n}\psi_{x,y,m}^{n}}}{{{c^2}{{\left\| {{\mathbf{p}_{n\left| {n - 1} \right.}} - {\mathbf{b}_m}} \right\|}^4}}}} + {\varsigma_{24}^n}}.
\end{align}

Next, ${\mathbf{J}_{{\mathbf{v}_n}{\mathbf{v}_n}}}$ can be expanded as ${\mathbf{J}_{{\mathbf{v}_n}{\mathbf{v}_n}}} = \left[ {\begin{array}{*{20}{c}}
{{J_{{v_x}{v_x}}^n}}&{{J_{{v_x}{v_y}}^n}}\\
{{J_{{v_x}{v_y}}^n}}&{{J_{{v_y}{v_y}}^n}}
\end{array}} \right]$.
Based on the derivations in (\ref{G_m}), (\ref{OP}), (\ref{JP}), and (\ref{J_O}), the elements of ${\mathbf{J}_{{\mathbf{v}_n}{\mathbf{v}_n}}}$ are derived as
\begin{align}
&{J_{{v_x}{v_x}}^n}= {\sum\limits_{m = 1}^M {\frac{{4{{{f_m^2}}}\sigma_{{\mu_m}}^{- 2}{{\left( {\ell_{x,m}^{n}} \right)}^2}}}{{{c^2}{{\left\| {{\mathbf{p}_{n\left| {n - 1} \right.}} - {\mathbf{b}_m}} \right\|}^2}}}} + {\varsigma_{33}^n}}, \\
&{J_{{v_x}{v_y}}^n}= {\sum\limits_{m = 1}^M {\frac{{4{{{f_m^2}}}\sigma_{{\mu_m}}^{- 2} \ell_{x,m}^{n} \ell_{y,m}^{n}}}{{{c^2}{{\left\| {{\mathbf{p}_{n\left| {n - 1} \right.}} - {\mathbf{b}_m}} \right\|}^2}}}} + {\varsigma_{34}^n}}, \\
&{J_{{v_y}{v_y}}^n} = {\sum\limits_{m = 1}^M {\frac{{4{{{f_m^2}}}\sigma_{{\mu_m}}^{- 2}{{\left( {\ell_{y,m}^{n}} \right)}^2}}}{{{c^2}{{\left\| {{\mathbf{p}_{n\left| {n - 1} \right.}} - {\mathbf{b}_m}} \right\|}^2}}}} + {\varsigma_{44}^n}}.   \label{a2}
\end{align}

\bibliographystyle{IEEEtran}
\bibliography{references}
\end{document}